\documentclass[twocolumn]{aastex63}
\usepackage[super]{nth}
\usepackage{savesym}
\savesymbol{tablenum}
\usepackage[T1]{fontenc}
\usepackage{inputenc}
\usepackage{multirow}
\restoresymbol{SIX}{tablenum}

\shorttitle{The Size and Structure of Globular Cluster Systems}
\shortauthors{Lim et al.}

\begin{document}

\title{The Next Generation Virgo Cluster Survey (NGVS). XXVII.\\ 
The Size and Structure of Globular Cluster Systems and their Connection to Dark Matter Halos}

\correspondingauthor{Sungsoon Lim}
\email{sslim00@gmail.com}

\author[0000-0002-5049-4390]{Sungsoon Lim}
\affiliation{Department of Astronomy, Yonsei University, 50 Yonsei-ro Seodaemun-gu, Seoul, 03722, Republic of Korea}

\author[0000-0002-2073-2781]{Eric W. Peng}
\affiliation{NSF’s NOIRLab, 950 N. Cherry Avenue, Tucson, AZ 85719, USA}

\author[0000-0003-1184-8114]{Patrick C{\^o}t{\'e}}
\affiliation{Herzberg Astronomy and Astrophysics Research Centre, National Research Council of Canada, Victoria, BC V9E 2E7, Canada}

\author[0000-0002-8224-1128]{Laura Ferrarese}
\affiliation{Herzberg Astronomy and Astrophysics Research Centre, National Research Council of Canada, Victoria, BC V9E 2E7, Canada}

\author[0000-0002-0363-4266]{Joel C. Roediger}
\affiliation{Herzberg Astronomy and Astrophysics Research Centre, National Research Council of Canada, Victoria, BC V9E 2E7, Canada}

\author[0000-0002-4718-3428]{Chengze Liu}
\affiliation{Department of Astronomy, School of Physics and Astronomy, and Shanghai Key Laboratory for Particle Physics and Cosmology, Shanghai Jiao Tong University, Shanghai 200240, China}

\author[0000-0002-1685-4284]{Chelsea Spengler}
\affiliation{Institute of Astrophysics, Pontificia Universidad Cat\'olica de Chile, Av. Vicu\~{n}a Mackenna 4860, 780436 Macul, Santiago, Chile}

\author[0000-0002-2814-3578]{Elisabeth Sola}
\affiliation{Universit{\'e} de Strasbourg, CNRS, 11 rue de l’Universit{\'e}, Strasbourg, France}

\author[0000-0003-3343-6284]{Pierre-Alain Duc}
\affiliation{Observatoire astronomique de Strasbourg, Universit{\'e} de Strasbourg, CNRS, UMR 7550, 11 rue de l’Universit{\'e}, F-67000 Strasbourg, France}

\author[0000-0002-3790-720X]{Laura V. Sales}
\affiliation{Department of Physics and Astronomy, 900 University Avenue, Riverside, CA 92521, USA)}

\author[0000-0002-5213-3548]{John P. Blakeslee}
\affiliation{NSF’s NOIRLab, 950 N. Cherry Avenue, Tucson, AZ 85719, USA}

\author[0000-0002-3263-8645]{Jean-Charles Cuillandre}
\affiliation{AIM Paris Saclay, CNRS/INSU, CEA/Irfu, Université Paris Diderot, Orme des Merisiers, F-91191 Gif-sur-Yvette Cedex, France}

\author[0000-0001-9427-3373]{Patrick R. Durrell}
\affiliation{Department of Physics and Astronomy, Youngstown State University, One University Plaza, Youngstown, OH 44555, USA}

\author[0000-0002-6155-7166]{Eric Emsellem}
\affiliation{European Southern Observatory, Karl-Schwarzschild Stra{\ss}e 2, D-85748 Garching bei M\"{u}nchen, Germany}

\author[0000-0001-8221-8406]{Stephen D. J. Gwyn}
\affiliation{Herzberg Astronomy and Astrophysics Research Centre, National Research Council of Canada, Victoria, BC V9E 2E7, Canada}

\author[0000-0002-7214-8296]{Ariane Lan{\c c}on}
\affiliation{Universit{\'e} de Strasbourg, CNRS, Observatoire Astronomique de Strasbourg, UMR 7550, 11 rue de l'Universit{\'e}, F-67000 Strasbourg, France}

\author[0000-0002-1442-2947]{Francine R. Marleau}
\affiliation{Institut f{\"u}r Astro- und Teilchenphysik, Universit{\"a}t Innsbruck, Technikerstra{\ss}e 25/8, Innsbruck, A-6020, Austria}

\author[0000-0002-7089-8616]{J. Christopher Mihos}
\affiliation{Department of Astronomy, Case Western Reserve University, Cleveland, OH 44106, USA}

\author[0000-0003-4552-9808]{Oliver M\"uller}
\affiliation{Institute of Physics, Laboratory of Astrophysics, École Polytechnique Fédérale de Lausanne (EPFL), 1290 Sauverny, Switzerland}

\author[0000-0003-0350-7061]{Thomas H. Puzia}
\affiliation{Institute of Astrophysics, Pontificia Universidad Cat{\'o}lica de Chile, Av. Vicu{\~n}a Mackenna 4860, 7820436 Macul, Santiago, Chile}

\author[0000-0003-4945-0056]{Rub{\'e}n S{\'a}nchez-Janssen}
\affiliation{STFC UK Astronomy Technology Centre, Royal Observatory, Blackford Hill, Edinburgh, EH9 3HJ, UK}

\begin{abstract}
We study the size and structure of globular clusters (GC) systems of 118 early-type galaxies from the NGVS, MATLAS, and ACSVCS surveys. Fitting S\'ersic profiles, we investigate the relationship between effective radii of GC systems ($R_{e, \rm gc}$) and galaxy properties. GC systems are 2--4 times more extended than host galaxies across the entire stellar mass range of our sample ($10^{8.3} < M_* < 10^{11.6}~M_{\odot}$). The relationship between $R_{e, \rm gc}$ and galaxy stellar mass exhibits a characteristic ``knee'' at a stellar mass of $M_p \simeq 10^{10.8}$, similar to galaxy $R_e$--stellar mass relationship. We present a new characterization of the traditional blue and red GC color sub-populations, describing them with respect to host galaxy $(g'-i')$ color ($\Delta_{gi}$): GCs with similar colors to their hosts have a ``red'' $\Delta_{gi}$, and those significantly bluer GCs have a ``blue'' $\Delta_{gi}$. The GC populations with red $\Delta_{gi}$, even in dwarf galaxies, are twice as extended as the stars, suggesting that formation or survival mechanisms favor the outer regions. We find a tight correlation between $R_{e, \rm gc}$ and the total number of GCs, with intrinsic scatter $\lesssim 0.1$ dex spanning two and three orders of magnitude in size and number, respectively. This holds for both red and blue subpopulations, albeit with different slopes.  Assuming that $N_{GC, Total}$ correlates with $M_{200}$, we find that the red GC systems have effective radii of roughly 1-5\% $R_{\rm 200}$, while the blue GC systems in massive galaxies can have sizes as large as $\sim$10\% $R_{\rm 200}$. Environmental dependence on $R_{e, \rm gc}$ is also found, with lower density environments exhibiting more extended GC systems at fixed mass. 
\end{abstract}


%%%%%%%%%%%%%%%%%%%%%%%%%%%%%%%%%%%%%%%%%%%%%%%%%%%%%%%%%%%%%%%%%%%%%%%%%%%%%%%%%%%%%%
%%%%%%%%%%%%%%%%%%%%%%%%%%%%%%%%%%%%%%%%%%%%%%%%%%%%%%%%%%%%%%%%%%%%%%%%%%%%%%%%%%%%%%
%%%
%%% INTRODUCTION (Section 1)
%%%
%%%%%%%%%%%%%%%%%%%%%%%%%%%%%%%%%%%%%%%%%%%%%%%%%%%%%%%%%%%%%%%%%%%%%%%%%%%%%%%%%%%%%%
%%%%%%%%%%%%%%%%%%%%%%%%%%%%%%%%%%%%%%%%%%%%%%%%%%%%%%%%%%%%%%%%%%%%%%%%%%%%%%%%%%%%%%

\section{Introduction}
\label{intro}

The size of a galaxy, i.e., the spatial extent of its baryons, is the result of a number of important physical processes that govern its evolution: angular momentum \citep{Mo98}, dissipation, and merging. Dissipational star formation, possibly the result of wet mergers or interactions, can reduce the size of a galaxy by driving gas to the center and preferentially forming stars there \citep{Tac16}. By contrast, dissipationless interactions like dry mergers and tidal heating, can increase the sizes of galaxies without adding very much stellar mass \citep{Naa09}. As a result the effort to measure the sizes of galaxies as a function of their stellar mass, halo mass, and other fundamental properties as a function of redshift, as well as reproduce those sizes in simulations \citep[e.g.,][]{Fur17,Gen18}, has been central to our efforts to understand galaxy evolution. 

While it may be possible to have a single number for a galaxy's ``size'' (usually the effective or half-light radius), studies of nearby galaxies have long shown that galaxies in fact have multiple distinct structural components (disks, bulges, bars, halos), each of which has its own characteristic size scale, and which reflects a different phase of evolution. Galaxy stellar halos are the most obvious example of how different structural components in a galaxy can have quite different spatial distributions and evolutionary histories. Stellar halos are the most extended and diffuse structures in galaxies, and yet also contain a stellar population that represents some of the most intense star formation in a galaxy's evolution: globular clusters. This apparent contradiction, as well as many other properties, makes globular clusters, and their apparently large spatial extent relative to their host galaxies, an intriguing target for study.

Globular clusters (GCs) --- compact, dense associations of stars that surround galaxies spanning wide ranges in mass, type and environment --- are valuable probes of galaxy formation processes. Similar to the stars that make up the bulk of their host galaxies, the ubiquity of GCs requires any complete theory of galaxy formation to explain not just the {\it existence} of GC systems but also their detailed properties, including the observed trends with host galaxies properties. Compared to the stars that make up galaxies, GCs offer some distinct advantages as galaxy formation probes. From a purely observational perspective, their compact nature ($\langle{r_h}\rangle \simeq 3$ pc) and high luminosities ($\langle{L}\rangle \simeq 10^5$~$L_\odot$) allow them to be readily recognized against the diffuse stellar light of the underlying galaxy. As Nature's closest approximations to simple stellar populations (SSPs), it is possible from imaging and spectroscopy to deduce ages and abundances for individual clusters, and thereby investigate the past history of star formation and chemical enrichment in their host galaxies \citep[e.g.,][and references therein]{Bro06}. If radial velocity measurements are also available, then GCs can  be used as dynamical probes for studying the amount and spatial distribution of gravitating mass within galaxies --- from their centers to their outermost regions, where dark matter dominates \citep[e.g.,][]{Ala17,Tol18,Fen19,Mul20}.

Over the past two decades, our understanding of the detailed properties of GC systems increased dramatically, thanks mainly improvements in observing facilities. For example, the deployment of the ACS and WFC3 cameras on the Hubble Space Telescope (HST) made it possible to obtain high-resolution imaging across the ultraviolet-optical-infrared region for GCs belonging to many nearby galaxies, including both dwarfs and giants \citep[e.g.,][]{Cot04,Jor07}. On the ground, wide-field imaging performed with 2.5 to 4m-class telescopes, and multi-object spectroscopy with  8-10-m-class telescopes, have also led to significant improvements in our understanding of the properties of GC systems, especially those associated with high- and intermediate-luminosity galaxies: i.e., their overall numbers, formation efficiencies, luminosity and mass functions, color distributions, color-magnitude relations, and dynamical properties \citep[see, e.g., the review of][]{Bro06}.

On the other hand, our knowledge of {\it spatial distributions} for GC systems remains far more limited. This issue has taken on a renewed importance in recent years, for several reasons. First, cosmological simulations have reached a level of sophistication where it is now possible to make quantitative predictions for the spatial distributions of GC systems and their dependence on galaxy mass and assembly history \citep{Kru15,Kru19,Kru20,Rei19,Rei21,Che22}. Such predictions must be carefully tested against observations for GCs in real galaxies, ideally selected to span wide ranges in mass, type and environment. Meanwhile, an increasing number of studies have used data from the literature to explore the connection between the total numbers of GCs in galaxies and the underlying mass in baryons and dark matter --- with the latter quantity usually inferred from an assumed stellar-to-halo mass relation \citep{Pen08,Hud14,Har15,Har17,For16,Hud18,For18}. These studies suggest that GC systems contribute a nearly constant fraction of the total dark matter mass, supporting earlier claims based on observations of GCs in massive cluster galaxies (e.g., \citealt{Bla97,Bla99,McL99}). 

From a practical perspective, measurements of the size and the richness of GC systems are linked in a fundamental way: i.e., a reliable determination of the total number of GCs associated with a galaxy generally requires a measurement of its GC density profile, which can be integrated to give the total number of GCs. In principle, this is straightforward, but our knowledge of GC density distributions, and their variation along the mass spectrum of galaxies, remains surprisingly limited. Most  investigations into the structure of GC systems either combined photographic and CCD data for a few bright galaxies \citep{Har86,McL94}, performed GC counts using mosaic CCD cameras on ground-based telescopes for small samples of intermediate-mass galaxies based on GC counts (e.g. \citealp{Rho01,Rho04,Rho07,Har12,Har14,Ko19}). Several studies have examined the spatial distribution of GCs using larger sample sizes. \citep{Kar14} proposed a linear relationship between the effective radius of galaxies and the effective radius of GC systems based on six galaxies from the SLUGGS survey \citep{Bro14}. \citet{Zar15} studied the GC properties of 97 early-type galaxies using S$^4$G survey images \citep{She10}, and they estimated the total number of GCs by fitting the GC radial number densities with a power-law function with a fixed power value. \citet{Hud18} and \citet{For17} studied GC spatial distributions using $20--30$ galaxies, and suggested the linear relations between galaxy stellar mass and GC system effective radii, but slopes of the relations in two studies are significantly different. \citet{Cas19} and \citet{DeB22} showed different slopes for the relationships between galaxy stellar mass and effective radius of GC systems, distinguishing between high and low mass galaxies. A power-law relationship between the effective radius of GC systems and the total number of GCs has been suggested. However, these studies are limited by the small number of galaxies considered; or by the use of GC counts from narrow-field HST images, which often require substantial, sometimes uncertain, corrections to account for missed GCs at larger radii; or by the use of more assumptions about the GC radial distribution, which provide limited information about the GC spatial distribution.

In this paper, we build on this latter approach by presenting a comprehensive study of the structure and sizes of the GC systems associated with more than one hundred galaxies in the local Universe. Our analysis relies on imaging for 87 early-type galaxies in the Virgo Cluster acquired as part of the {\it Next Generation Virgo Cluster Survey} 
(NGVS, \citealp{Fer12, Fer20}) --- a multi-band imaging survey undertaken with the Canada-France-Hawaii Telescope (CFHT). The depth, areal coverage and 
multi-band nature of the NGVS\footnote{https://www.ngvs-astro.org/} makes it an ideal resource for studying GCs in nearby galaxies. Previous papers in the NGVS series 
have examined: the distribution of GCs within galaxies and across the entirety of the cluster (\citealt{Mun14,Dur14,San19,Lim20}); the stellar populations within 
GCs (\citealt{Pow16a,Pow16b,Pow17,Pow18,Ko21}); and galaxy kinematics and dynamics across a range of spatial scales using GC radial velocities (\citealt{Zhu14,Zha15,Tol16,Tol18,Lon18,Li20,Tay21}). For 75 of the galaxies in our sample, HST imaging is also available from the ACS Virgo Cluster 
Survey\footnote{https://www.acsvcs.org/} \citep{Cot04}, 
which allows us to accurately measure GC density profiles into the core region. To supplement the NGVS sample, we include 31 additional galaxies 
from the {\it Mass Assembly of early-Type GaLAxies with their fine Structures} (MATLAS, \citealp{Duc15,Duc20}) survey. Like the
NGVS, the multi-band MATLAS\footnote{http://obas-matlas.u-strasbg.fr/WP/} survey was carried out with CFHT and has the benefit of depth, wide field coverage
and sub-arcsecond image quality.

This paper is structured as follows. In \S2, we describe the data analyzed in this paper and the methodology used to study the spatial distribution of GCs in our sample galaxies.  In \S3, we present our determinations of the effective radii of our GC systems and show how the measured sizes correlate with other galaxy parameters, including stellar masses, stellar radii and halo masses. In \S4, we discuss the implications of our results within the context of current models of galaxy formation. In \S5, we summarize our findings and outline some directions for future work.

%%%%%%%%%%%%%%%%%%%%%%%%%%%%%%%%%%%%%%%%%%%%%%%%%%%%%%%%%%%%%%%%%%%%%%%%%%%%%%%%%%%%%%
%%%%%%%%%%%%%%%%%%%%%%%%%%%%%%%%%%%%%%%%%%%%%%%%%%%%%%%%%%%%%%%%%%%%%%%%%%%%%%%%%%%%%%
%%%
%%% OBSERVATIONS (Section 2) - Table 1, Figures 1-9
%%%
%%%%%%%%%%%%%%%%%%%%%%%%%%%%%%%%%%%%%%%%%%%%%%%%%%%%%%%%%%%%%%%%%%%%%%%%%%%%%%%%%%%%%%
%%%%%%%%%%%%%%%%%%%%%%%%%%%%%%%%%%%%%%%%%%%%%%%%%%%%%%%%%%%%%%%%%%%%%%%%%%%%%%%%%%%%%%

\section{Observations, Data and Methods} 
\label{data}

Our analysis is based on data acquired in two complementary imaging surveys undertaken with MegaCam on the Canada-France-Hawaii Telescope and supplemented by imaging from the ACS/WFC instrument on HST. Table~\ref{tab:surveys} summarizes the data used in this paper, including survey names and filters, numbers of galaxies, and the average values and full ranges spanned by our respective samples in magnitude, color and stellar mass.

\subsection{Surveys and Sample Definition}
\label{data:surveys}

\begin{deluxetable*}{lcccr}
\tablecaption{Sample galaxies and their properties. 
\label{tab:surveys}}
\tablehead{
\colhead{Survey} &\colhead{$N$} & \colhead{$\langle{M_g}\rangle$ [max,min]} & \colhead{$\langle{(g-i)_0}\rangle$ [min,max]} & \colhead{$\langle{\log_{10}{M_{*}/M_{\odot}}}\rangle$ [min,max]}} %\\
\startdata
NGVS ($u^*g'i'z'$) & $87$ & $-18.6$ [$-15.5,-22.6$] & $0.95$ [0.60,1.15] & $9.93$ [8.31,11.56] \\
MATLAS ($u^*g'r'i'$) & $31$  & $-19.8$ [$-18.2,-21.2$] & $0.94$ [0.78,1.06] & $10.51$ [9.60,11.43]\\
ACSVCS ($g_{475}$ $z_{850}$)  & $75^a$ & $-18.4$ [$-15.5,-22.6$] & $0.94$ [0.60,1.08] & $9.85$ [8.31,11.56] \\
\hline
Full sample  & $118$  & $-18.9$ [$-15.5,-22.6$] & 0.95 [0.60,1.15] & 10.08 [8.31,11.56]\\
\enddata
\tablecomments{$^a$ACSVCS data is supplementary for the NGVS, so these galaxies are included in the NGVS. Full list of galaxies is available in Table \ref{tbl:galaxies}}

\end{deluxetable*}

The primary survey used in this analysis is the NGVS, a $u^*g'i'z'$ imaging survey of the Virgo cluster carried out with Megacam on CFHT \citep{Bou03}. The survey covers an area of 104 deg$^2$ --- from the cluster core out to the virial radii of its two main sub-clusters, Virgo A and B. The survey is comprised of 117 distinct pointings, with a uniform (point-source) limiting magnitude of 25.9~mag, and a median seeing of $0\farcs80$ in the $g$-band. Full details on the survey, including observing and data processing strategies, are available in \cite{Fer12,Fer20}. 

We also rely on data from MATLAS \citep{Duc15,Bil22}, a deep imaging survey of ATLAS$^{\rm 3D}$ galaxies \citep{Cap11} that is also based on imaging from CFHT Megacam. The ATLAS$^{\rm 3D}$ survey targeted 260 massive early-type galaxies with distances $d <42$ Mpc and magnitudes $M_K  < -21.5$ mag \citep{Cap11}. There are 58 ATLAS$^{\rm 3D}$ galaxies within the NGVS footprint, and the MATLAS survey uses NGVS data for these objects. Outside of the NGVS region, the MATLAS survey was designed to be carried out with the $u^*g'r'i'$ filters but, in practice, most galaxies were observed in a subset of these filters ($g'$ and $r'$ being most common). For this study, we consider only the subset of MATLAS galaxies with MegaCam photometry in at least three passbands, usually $g'r'i'$ or $u^*g'r'$. The seeing varies with targets and filters but, on average, image quality in the $g$-band is $\simeq 0\farcs84$. The details for dithering strategies and stacking images are described in \citet{Duc15}. The MegaPipe data reduction and processing pipeline \citep{Gwy08} provides the astrometric and photometric calibration of GC candidates. The surveys have a similar limiting surface brightness of $\mu_g\simeq29.0$ mag ${\rm arcsec^{-2}}$, but limiting magnitudes can vary significantly depending on the seeing for specific targets.

%%%%%%%%%%%%%%%%%%%%%%%%%%%%%%%%%%%%%%%%%%%%%%%%%%%%%%%%
\begin{figure}
\epsscale{1.15}
\plotone{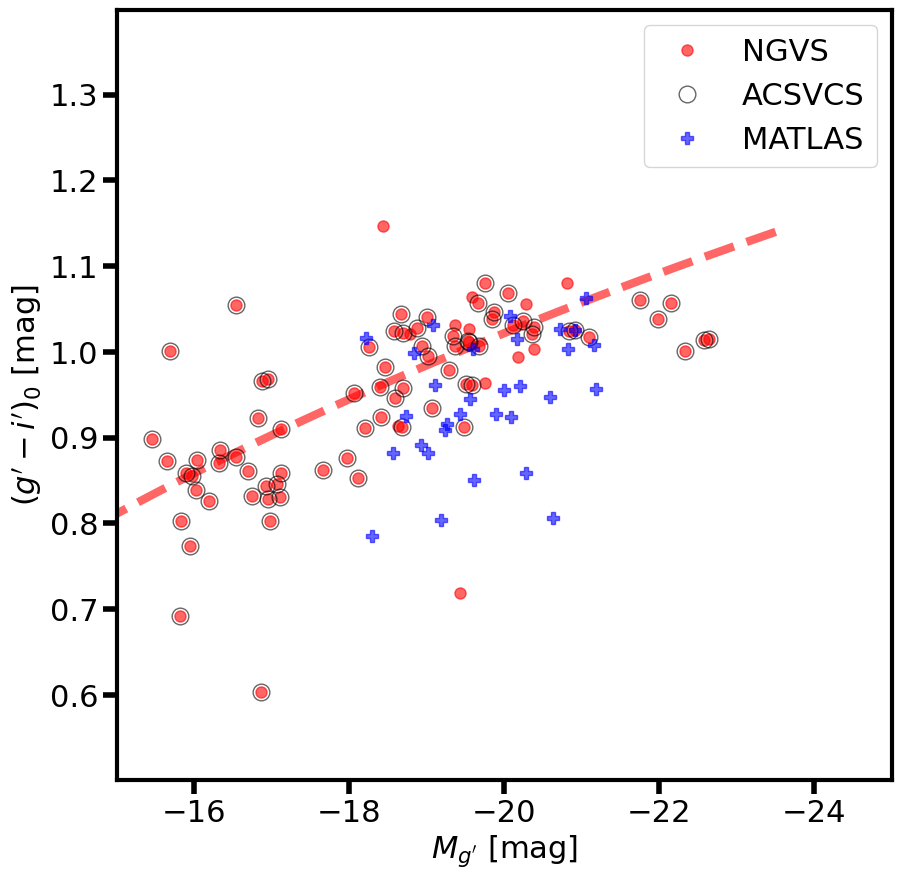}
\caption{Color-magnitude diagram for our program galaxies. Red filled circles, black open circles, and blue crosses show galaxies from NGVS, ACSVCS and MATLAS, respectively. The dashed curve shows the red sequence defined by $\sim$400 galaxies in the  Virgo cluster core \citep{Roe17}. 
\label{fig:cmd}}
\end{figure}
%%%%%%%%%%%%%%%%%%%%%%%%%%%%%%%%%%%%%%%%%%%%%%%%%%%%%%%%

Among the NGVS and MATLAS samples, we have chosen galaxies for our analysis in the following way. First, we selected all MATLAS galaxies (including the 58 in the NGVS region) observed in three or more filters. Because GCs at distances of $\sim10$--$45$ Mpc appear mostly as point sources in ground-based images, we distinguish GCs from foreground stars using their colors. Although we could select GC candidates using magnitudes and a single color index, we have used color-color diagrams to decrease contamination by foreground stars and compact, background galaxies. Second, we limited our targets to distances of 25 Mpc or less in order to detect enough GCs for an analysis of their spatial distribution. Finally, we included a sample of dwarf galaxies in the NGVS that were also observed in ACS Virgo Cluster Survey \citep{Cot04}, where images taken with the ACS instrument \citep{For98} in WFC mode can be used to select GCs with high confidence \citep[see, e.g.,][]{Jor04,Jor09}. We have 139 galaxies with the above criteria as a parent sample. As described in \S2.3, not all of these galaxies yielded reliable GC radial number density profile fits, and so from a parent sample of 139 galaxies, the final sample used in our analysis consists of 118 galaxies. Of these, 87 galaxies are from the NGVS, with 75 of them having ACSVCS data. Additionally, 31 galaxies are from the MATLAS. Given that the NGVS and MATLAS surveys have data in common, to reduce confusion we refer to "MATLAS galaxies" as those MATLAS galaxies outside of Virgo, i.e., not in the NGVS footprint. 

%%%%%%%%%%%%%%%%%%%%%%%%%%%%%%%%%%%%%%%%%%%%%%%%%%%%%%%%
\begin{figure}
\epsscale{1.15}
\plotone{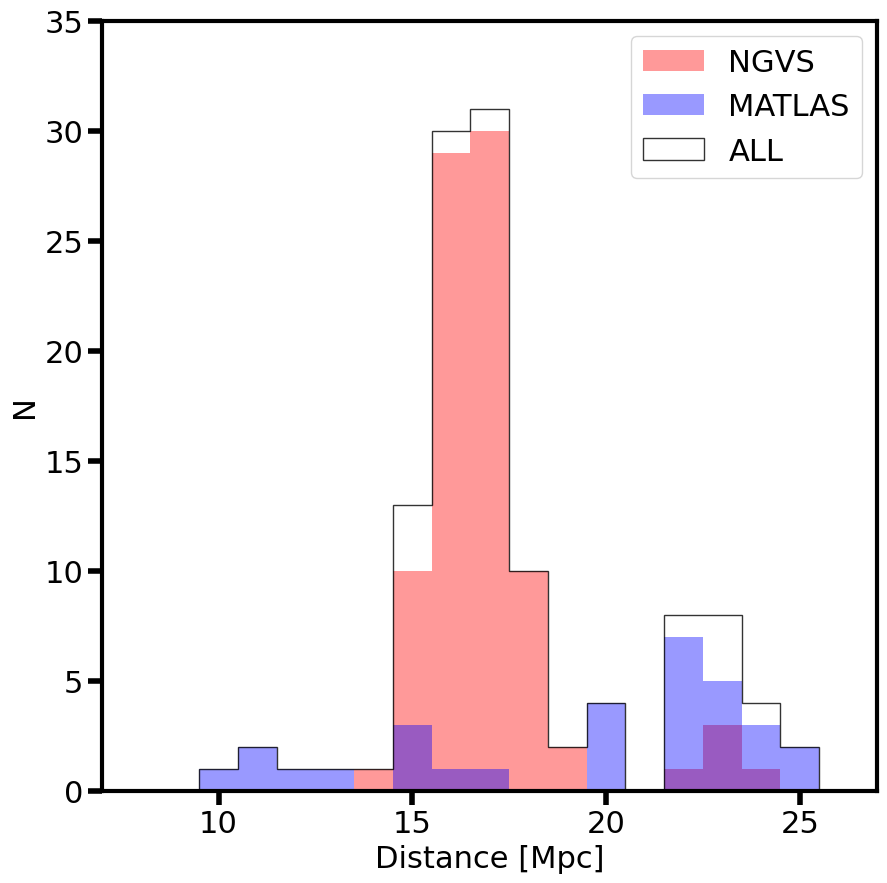}
\caption{The distribution of distances for galaxies in this study. Red and blue histograms show results for the NGVS and MATLAS samples, respectively. The black solid histgram shows results for total samples in this study. The MATLAS galaxies in this study have distances $d \lesssim 25$ Mpc. For ACSVCS galaxies, distances are based on HST surface brightness fluctuation measurements \citep{Mei07,Bla09,Can18}, while distances for MATLAS and NGVS (not in ACSVCS) galaxies are taken from \citet{Cap11}. These are stacked histograms, so the top edge line shows the distribution of distances for the entire sample. 
\label{fig:distancehist}}
\end{figure}
%%%%%%%%%%%%%%%%%%%%%%%%%%%%%%%%%%%%%%%%%%%%%%%%%%%%%%%%

\subsection{Sample characteristics}

Figure~\ref{fig:cmd} presents a color-magnitude diagram for our program galaxies. All magnitudes and colors of galaxies are estimated on NGVS and MATLAS images. As expected, our targets roughly follow the red sequence of galaxies in the central region of Virgo cluster \citep{Roe17}. 
The galaxies in the MATLAS survey also lie on the red sequence but seem to have a little larger scatter than those in the NGVS survey.
Figure~\ref{fig:distancehist} are the distribution of galaxy distances. All galaxies are more distant than 10 Mpc, with the great majority of our targets belonging to the Virgo Cluster at $d\simeq16.5$~Mpc \citep{Mei07}. Our sample includes a handful of galaxies beyond 20 Mpc --- some from the MATLAS program, and some from the NGVS, which includes a few members of the W$^\prime$ Cloud at $d\sim23$~Mpc \citep{Mei07,Bla09}. 

%%%%%%%%%%%%%%%%%%%%%%%%%%%%%%%%%%%%%%%%%%%%%%%%%%%%%%%%
\begin{figure}
\epsscale{1.15}
\plotone{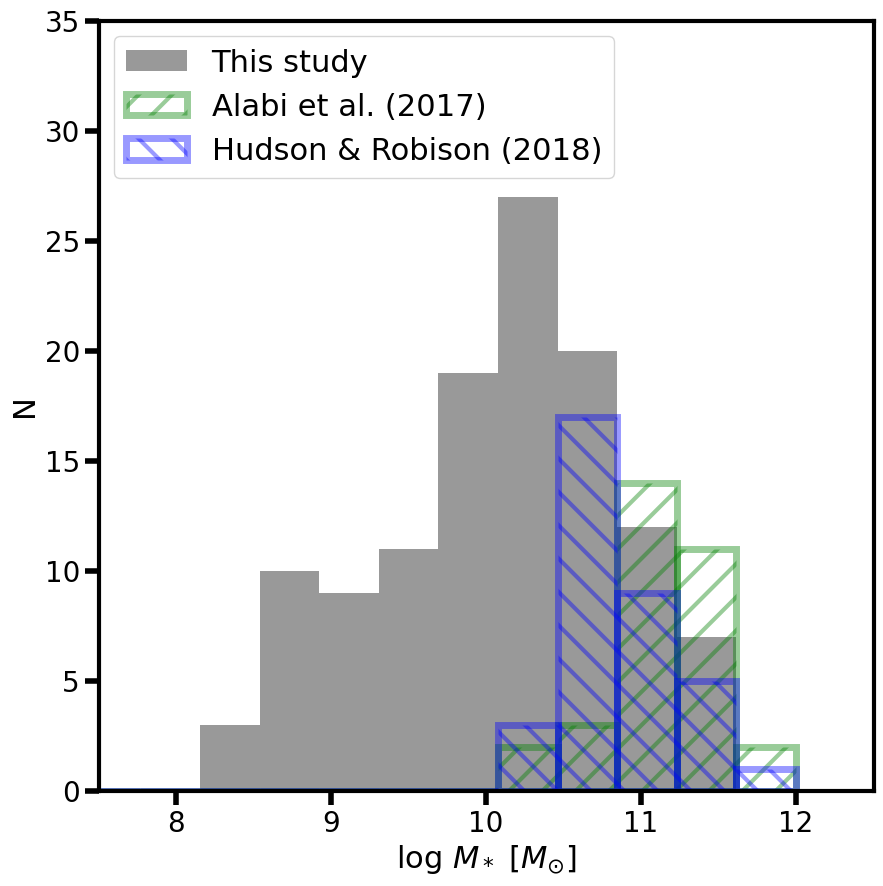}
\caption{Distribution of stellar masses, $M_*$, for our program galaxies (gray histogram). For comparison, the green and blue hatched histograms show the distribution of stellar masses for the samples of \cite{Ala17} and \cite{Hud18}. There are 32 and 35 galaxies in \citet{Ala17} and \citet{Hud18}, respectively.
\label{fig:samplehist}}
\end{figure}
%%%%%%%%%%%%%%%%%%%%%%%%%%%%%%%%%%%%%%%%%%%%%%%%%%%%%%%%

Stellar masses for our program galaxies were calculated by modelling of their spectral energy distributions (SEDs) in all available bands (i.e., as many of $u^*g'r'i'z'$ as were available). We used the Flexible Stellar Population Synthesis models of \cite{Con09} and assumed exponentially declining star formation histories and a Chabrier initial mass function. The SEDs were fitted to a grid of 50,000 synthetic models with metallicities in the $0.01 \le Z/Z_{\odot} \le 1.6$ range, star formation timescales $0.5 \le \tau \le 100$ Gyr$^{-1}$, and luminosity-weighted ages between 5 and 13 Gyr. The mass errors are derived from the 16th and 84th percentiles of the marginalized posterior for the mass: we obtain samples of the posterior from our MCMC analysis, find the 16th and 84th percentiles, and set the error to $0.5 \times$ (P84 - P16). Mean mass error of our sample is about 0.04 in log with maximum and minimum of 0.10 and 0.01, respectively. All details of stellar mass estimation are in Roediger et al. (2023, in prep.). Figure~\ref{fig:samplehist} shows the distribution of these stellar masses. For comparison, we also show in this figure stellar mass distributions for galaxies targeted in two recent studies of GC systems \citep{Ala17,Hud18}. Compared to these previous studies, our program galaxies span a significantly wider range in stellar mass (a factor of roughly $\sim$2000 with the faintest dwarfs in our program having stellar masses of $M_* \simeq 2\times10^8M_{\odot}$. 

%%%%%%%%%%%%%%%%%%%%%%%%%%%%%%%%%%%%%%%%%%%%%%%%%%%%%%%%
\begin{figure}
\epsscale{1.15}
\plotone{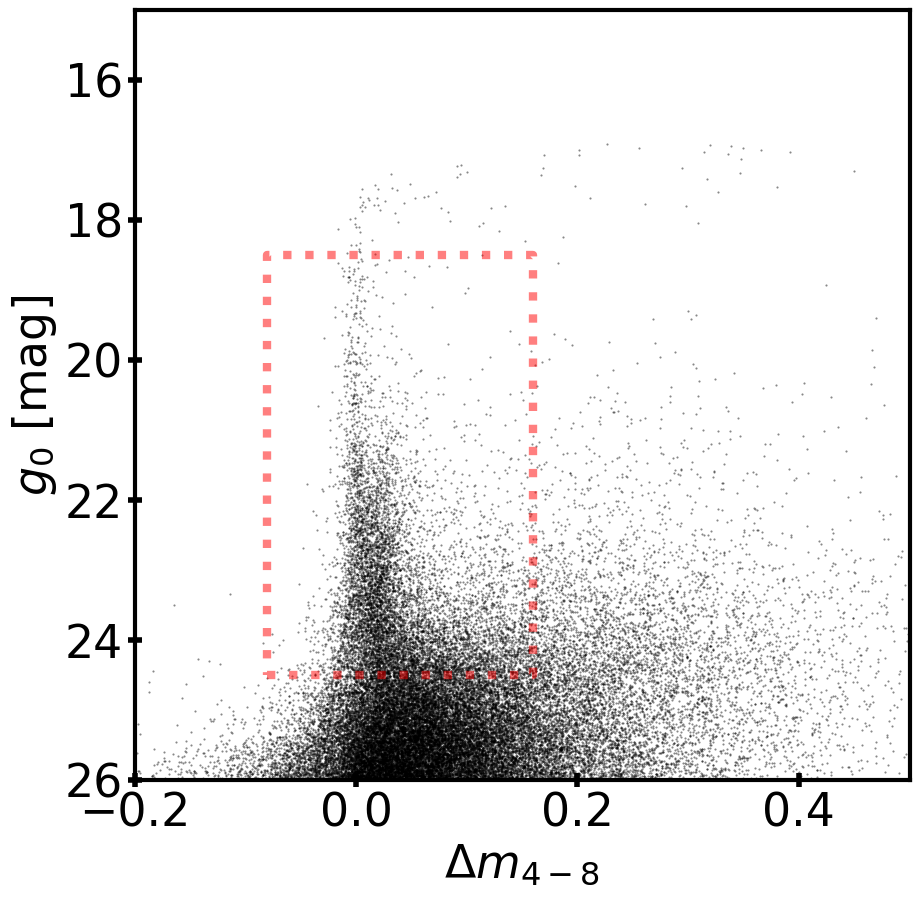}
\caption{Magnitude versus inverse concentration index, ${\rm \Delta m_{4-8}}$, for sources in the field of NGC4472 (M49, VCC1226), one of our NGVS program galaxies. The ${\rm \Delta m_{4-8}}$ index represents the magnitude difference between apertures with diameters of 4 and 8 pixels. Globular cluster candidates are selected to fall within the dotted region.
\label{fig:ic}}
\end{figure}
%%%%%%%%%%%%%%%%%%%%%%%%%%%%%%%%%%%%%%%%%%%%%%%%%%%%%%%%

\subsection{Photometry, Source Catalogs and Globular Cluster Selection}
\label{data:photometry}

Our methods for source detection and photometry are similar to those used in \citet{Dur14} and \citet{Liu15}. Briefly, we used Source Extractor \citep{Ber96} in dual-image mode to detect sources and perform photometric measurements. For NGVS target galaxies, we used the $g'$-band image as a detection image. Fluxes were measured within circular apertures for a variety of radii, and magnitudes corrected to $16$-pixel (3\arcsec) diameter aperture magnitudes. These magnitudes were then transformed onto the standard AB system by using PSF magnitudes of bright SDSS stars \citep{Alb17} within the field. A similar methodology was adopted for the MATLAS galaxies, although in this case,  $r'$-band images were used for source detection when the $g'$-band images suffered from significantly poorer seeing ($\gtrsim1.\arcsec5$).

%%%%%%%%%%%%%%%%%%%%%%%%%%%%%%%%%%%%%%%%%%%%%%%%%%%%%%%%
\begin{figure}
\epsscale{1.15}
\plotone{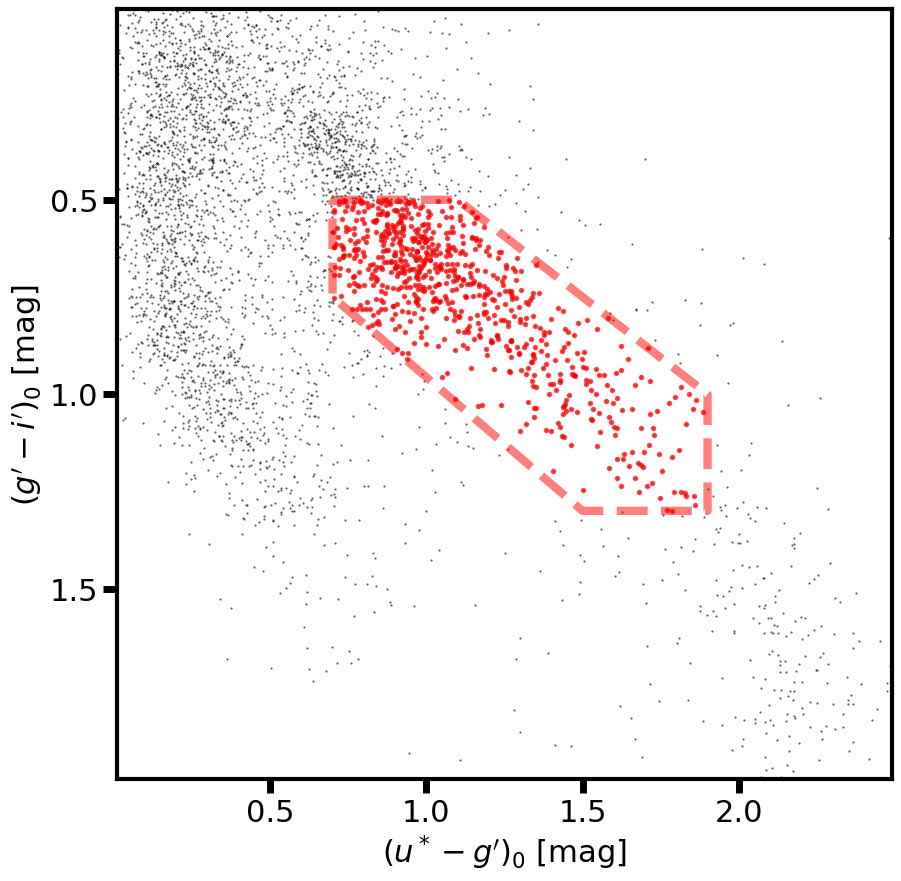}
\caption{$(u^*-g')_0-(g'-i')_0$ color-color diagram for point-like sources in the field of NGC4564 (VCC1664). Globular cluster candidates are selected using the red dashed polygon which is defined as the region occupied by spectroscopically confirmed globular clusters in NGC4486 (M87, VCC1361; \citealp{Mun14,Pow16b,Lim17,Ko21}). 
\label{fig:colorcolor}}
\end{figure}
%%%%%%%%%%%%%%%%%%%%%%%%%%%%%%%%%%%%%%%%%%%%%%%%%%%%%%%%

We subtracted the diffuse component of each galaxy from the original image to ensure reliable photometry for GCs located in the central regions of galaxies. 
For NGVS galaxies, we used customized two-dimensional models for diffuse light subtraction, as described in \citet{Fer20}. Galaxy models were fitted and subtracted independently in each band.  For the MATLAS galaxies, we used ring median models generated with the {\tt RMEDIAN} task in {\tt IRAF} \citep{Tod86,Tod93} to subtract the underlying galaxy light. Galaxy subtraction aside, all photometric measurements were carried out using the same methodology as for the NGVS targets.

Completeness tests were carried out using artificial stars generated with DAOPHOT \citep{Ste87}. We injected 5000 artificial stars into model subtracted images in each observing field, and derived the recovery rate as a function of both magnitude and underlying surface brightness.
Most regions have recovery rates in excess of $90\%$ for $g \lesssim 24.5$ mag, apart from the expected sharp decrease in the core regions of high surface brightnesses galaxies.

GC candidates were then selected on the basis of their sizes (i.e., compactness) and colors. 
Most GCs appear as point sources in our survey, so we selected point sources using a measure of how extended an object is relative to a point source (i.e., ``inverse concentrations''); these indices are the magnitude differences between 4-pixel (0\farcs75) and 8-pixel (1\farcs5) diameter aperture magnitudes, $\Delta m_{4-8}$, normalized so that point sources have a mean value of zero \citep{Dur14}. We chose sources with $-0.08<\Delta m_{4-8}<0.08$ as point sources.
This index is actually measured in two filters ($g'$ and $i'$ for NGVS and top two best seeing filters for MATLAS), where the final $\Delta m_{4-8}$ is the combination in quadrature of the measurement of each object in the two filters, i.e., $\Delta m_{4-8} = \sqrt{(\Delta g'_{4-8})^2 + (\Delta i'_{4-8})^2}$ for NGVS. 
Because GCs in nearby galaxies  ($d\lesssim$ 20 Mpc) can be slightly extended in images taken during conditions of good ground-based seeing ($\lesssim0.8$), we used a slightly expanded range, ($-0.08<\Delta m_{4-8}<0.16$, Figure~\ref{fig:ic}). We also required GC candidates to have $m_g <24.5$~mag because $\Delta m_{4-8}$ values for point sources and extended sources begin to merge below $g' \sim 24.5$~mag. 

From these ``point-like" sources, we selected GC candidates based on their colors. Using spectroscopically confirmed GCs in M87 as a guide, $u^*g'i'$ or $g'r'i'$ color-color diagrams were used to select likely GCs (see Figure~\ref{fig:colorcolor}). For M87, the polygons shown in the $u^*g'i'$ and $g'r'i'$ color-color diagrams include all spectroscopically confirmed GCs.

Finally, we used ACSVCS photometry \citep{Cot04} in the central regions to supplement the NGVS data for galaxies overlapped with ACSVCS galaxies in our NGVS sample. GC catalogs were taken from \citet{Jor09}, and transformed from the HST to CFHT MegaCam systems using relations fitted with photometry of GCs in M87. Followings are the fitted relations:
\begin{eqnarray}
\label{hstrelation}
m_{g',{\rm CFHT}} = m_{g, {\rm ACS}} -0.0574 \times (g-z)_{\rm ACS} -0.06 \\
(g'-i')_{\rm CFHT} = ((g-z)_{\rm ACS} + 0.27)/1.65
\end{eqnarray}
GC catalogs generated from these deep, high-resolution images are crucial for measuring the density profiles of GC systems in galaxy cores.

%%%%%%%%%%%%%%%%%%%%%%%%%%%%%%%%%%%%%%%%%%%%%%%%%%%%%%%%
\begin{figure}
\epsscale{1.15}
\plotone{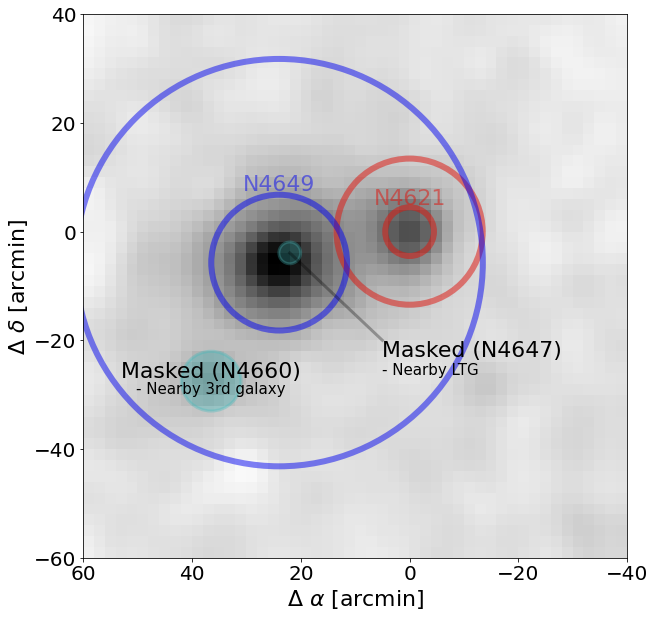}
\caption{Density map for GC candidates in the vicinity of two of our program galaxies, NGC4649 and NGC4621. The image measures $100\arcmin\times100\arcmin$, with a pixel size of $1\arcmin\times1\arcmin$, and has been smoothed using a 2-pixel Gaussian filter. North is up and East is to the left. Inner and outer circles are drawn at radii of $R_{e,{\rm gc}}$ and  $3R_{e,{\rm gc}}$ centered on each galaxy. The two GC systems have been fitted simultaneously in our analysis, while two other galaxies that are not members of our sample (NGC4660 and NGC4647) have been masked.
\label{fig:map}}
\end{figure}
%%%%%%%%%%%%%%%%%%%%%%%%%%%%%%%%%%%%%%%%%%%%%%%%%%%%%%%%

\subsection{Globular Cluster Density Profiles}
\label{data:fitting}

Our goal is to measure radial number density profiles for GC candidates in our program galaxies and use these profiles to explore the spatial distribution of the GC systems as a function of galaxy properties. An example of our methodology is presented in Figure~\ref{fig:map} which shows a Gaussian-smoothed density map for GC candidates in the vicinity of NGC4649 and NGC4621 --- two program galaxies that happen to fall within this single $100\arcmin\times100\arcmin$ field. As described below, we mask interloping galaxies and extract a radial number density profile for GC candidates associated with each program galaxy. Note that, unlike the case shown in Figure~\ref{fig:map}, the vast majority of our program galaxies are relatively isolated and can be fitted individually.

%%%%%%%%%%%%%%%%%%%%%%%%%%%%%%%%%%%%%%%%%%%%%%%%%%%%%%%%
\begin{figure}
\epsscale{1.15}
\plotone{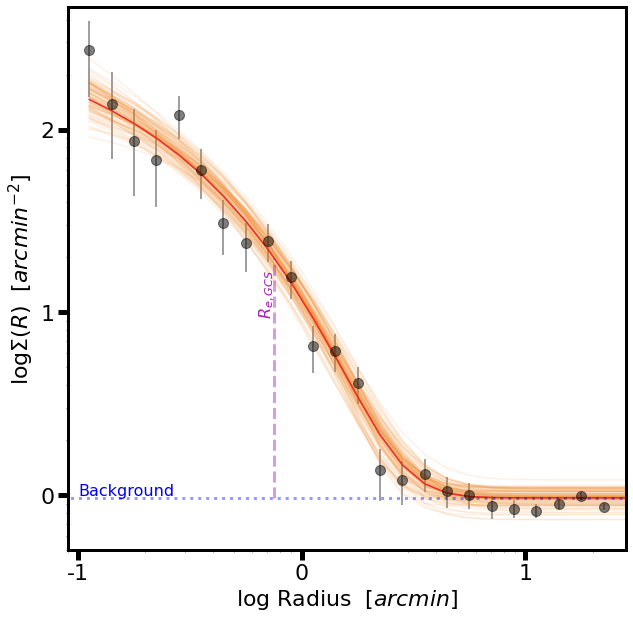}
\caption{Number density radial profile for GC candidates surrounding NGC4564. The filled circles show radial number densities in logarithmically spaced radial bins centered on the galaxy core. The red solid curve shows our best-fit S\'ersic (plus background) profile. The orange curves show 100 random resampling results that illustrate the range of fitting errors. The vertical dashed line and horizontal dotted line show the fitted effective radius of the GC system, $R_{e,{\rm gc}}$, and the underlying background level, $\Sigma_b$.
\label{fig:radprof}}
\end{figure}
%%%%%%%%%%%%%%%%%%%%%%%%%%%%%%%%%%%%%%%%%%%%%%%%%%%%%%%%

Figure~\ref{fig:radprof} shows a radial number density profile of GC candidates in NGC4564 (VCC1664), a typical galaxy for which both ACSVCS and NGVS imaging is available. For display purposes, we plot number densities computed in circular, logarithmically-spaced bins, although we rely on the locations of individual GC candidates in the profile fitting.
The density profiles are then parameterized with a modified S\'ersic function
\begin{equation}
\label{sersic}
\Sigma(R)=\Sigma_e  \exp \left\{ -b_n \left[ \left( \frac{R}{R_e} \right)^{1/n} -1 \right] \right\} + \Sigma_b
\end{equation}
\label{eq:sersic}
where 
\begin{eqnarray}
R & = & \sqrt{(X^{\prime 2} + Y^{\prime 2})/(1-\epsilon^2)} \\
X^{\prime} & = & (X-X_0)\cos{\theta} + (Y-Y_0)\sin{\theta} \\
Y^{\prime} & = & (Y-Y_0)\cos{\theta} - (X-X_0)\sin{\theta}.
\end{eqnarray}
Here $\theta$ is the position angle measured in the customary sense: i.e., from North to East. The fitted S\'ersic function provides us with an estimate for both the size ($R_e$) and concentration ($n$) of the GC system. Because the GC catalogs for each galaxy inevitably include some residual contamination (i.e., from foreground stars, background galaxies, and interloping GCs from nearby galaxies), we include a constant background term, $\Sigma_b$, in our fits. 

%%%%%%%%%%%%%%%%%%%%%%%%%%%%%%%%%%%%%%%%%%%%%%%%%%%%%%%%
\begin{figure}
\epsscale{1.15}
\plotone{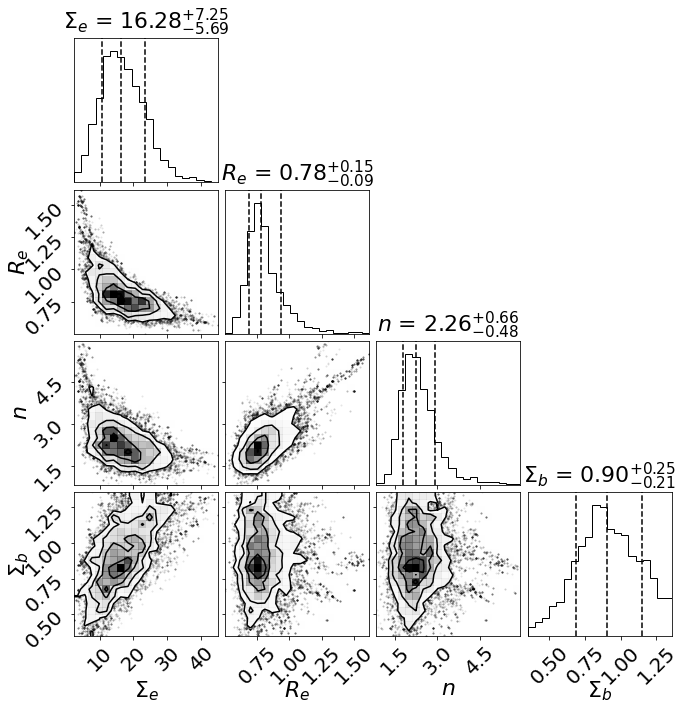}
\caption{Two-dimensional and marginalized posterior probability density functions for the effective radius ($R_e$), Sersic index ($n$), number density at the effective radius ($\Sigma_e$) and background level ($\Sigma_b$) for GC candidates in NGC4564. 
\label{fig:corner}}
\end{figure}
%%%%%%%%%%%%%%%%%%%%%%%%%%%%%%%%%%%%%%%%%%%%%%%%%%%%%%%%

%%%%%%%%%%%%%%%%%%%%%%%%%%%%%%%%%%%%%%%%%%%%%%%%%%%%%%%%
\begin{figure*}[htbp]
\centering
\includegraphics[width=0.99\textwidth]{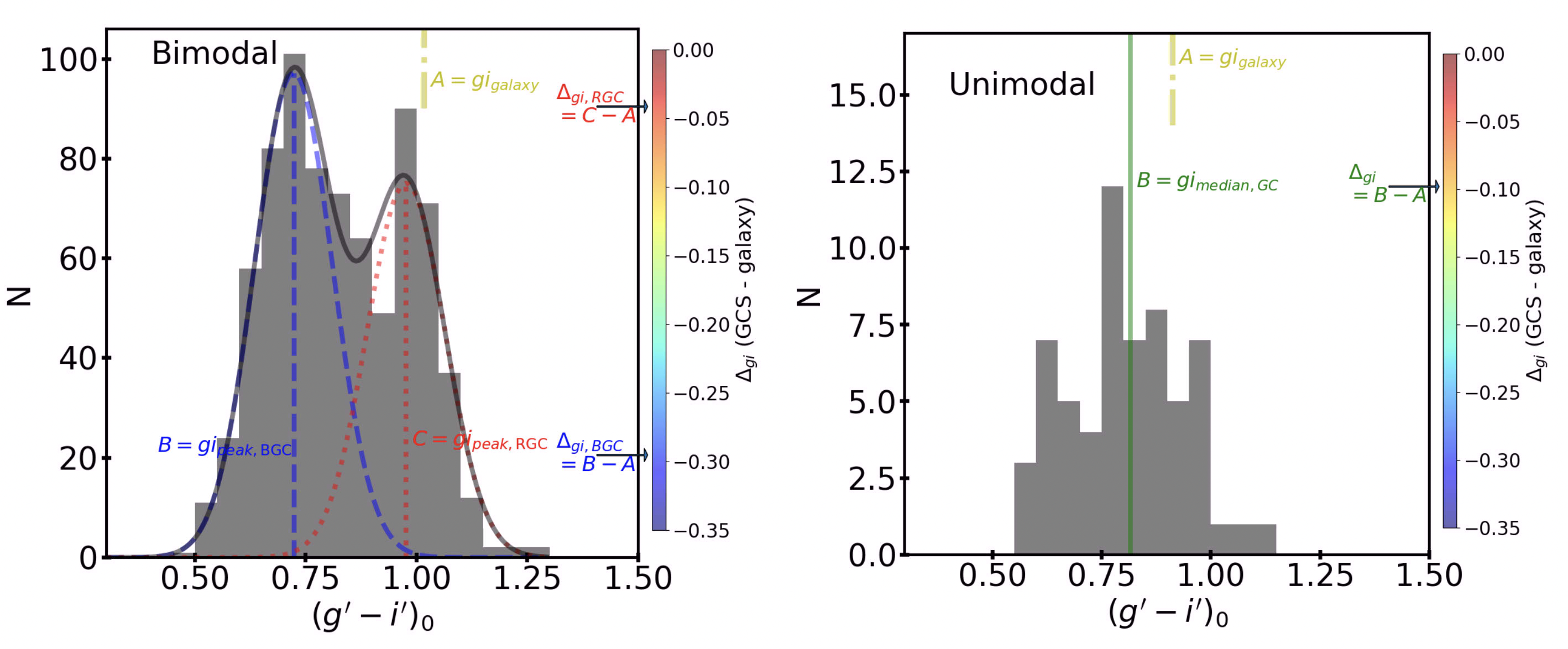}
\caption{GC color distributions for two of our sample galaxies which illustrate the approach used in this paper to analyze the color dependencies of GC system scaling relations. (Left panel). The GC color distribution for NGC4621, which is classified as bimodal by GMM \citep{Mur10}. In this case, the peak colors of the blue $(B)$ and red $(C)$ components are combined with the color of the underlying galaxy $(A)$ to define a color index for each of the blue and red GC subsystems: $\Delta_{gi,BGC}=(B-A)$ and $\Delta_{gi,RGC}=(C-A)$, respectively, and shown pointing to a corresponding position on the color bar at right. This color will be used in \S3. (Right panel) The GC color distribution for NGC4612, which is classified by GMM as unimodal. In this case, we define a single index $(B-A)$ as the difference between the color of the galaxy $(A)$ and the median color $(B)$ of its GC system.
\label{fig:colors}}
\end{figure*}
%%%%%%%%%%%%%%%%%%%%%%%%%%%%%%%%%%%%%%%%%%%%%%%%%%%%%%%%
%\color{blue}
This function was fitted to the data with a Markov Chain Monte Carlo (MCMC) method using the {\tt emcee} code in python \citep{For13}.  
Flat priors were set on several parameters: $\Sigma_e > 0.001$~arcmin$^{-2}$, $0.05 < n < 8.0$, and $0.05<R_e<30$ arcmin. 
For the ellipticity and position angle, we imposed flat priors of $0 \leq \epsilon < 0.1$ and $-10^{\circ} < \theta < 10^{\circ}$ due to difficulties in determining these values in the presence of GC contamination. For a handful of clearly elongated galaxies, we adopted a flat prior of $0 \leq \epsilon < 0.4$ and no prior for $\theta$.

For the background level, we adopted a Gaussian prior with mean and standard deviation estimated from the data. The mean of the Gaussian was estimated using the mean GC number density in a large annulus around the galaxy. For most fields, we used a ring with an inner radius of 30$^\prime$ and a width of 1$^\prime$. For the largest galaxies, an inner ring radius of 60$^\prime$ was used. The standard deviation of the Gaussian was estimated in the same annulus by dividing it into nine angular sub-sections (i.e., ``annular arcs''), and calculating the standard deviation of the mean backgrounds in these nine sub-annular regions. Using these values, we could adopt a reasonable Gaussian prior on the background level.

We employed a likelihood function as follows:
\begin{equation}
    \mathcal{L}(\Sigma_e,R_e,n,\Sigma_b) \propto \prod_i \ell_i (R_i | \Sigma_e,R_e,n,\Sigma_b)
\end{equation}
where $\ell_i (R_i | \Sigma_e,R_e,n,\Sigma_b)$ is the probability of finding the datum $i$ at radius $R_i$, given the three S\'ersic parameters, background level, and estimated completeness. To calculate the probability, we employed an areal integration of the modified S\'ersic function. While the standard S\'ersic function has a theoretical integrated functional form, this is not true of our modified S\'ersic function, so we used a numerical integration method to generate the likelihood function. 
The integration range was normally from the radius corresponding to the 50\% completeness limit to 30$^\prime$. For NGVS galaxies, where ACSVCS imaging is available, the integration was started at the galaxy center. As with the largest galaxies, the outer limit was taken to be 60$^\prime$.
Figure~\ref{fig:corner} shows representative corner plots from MCMC fitting of the GC system in NGC4564. Aside from the familiar correlation between $R_e$ and $\Sigma_e$, there are generally no strong correlations between any of the other parameters except for $\Sigma_e$ and $\Sigma_b$. In a few cases, there are also weak correlations between S\'ersic index, $n$, and effective radius, $R_e$.\ In general, the fitted models provide good matches to the measured profiles, as shown in Figure~\ref{fig:radprof}.

As a final check on our results, we visually inspected the fitted profiles for each our program galaxies and excluded galaxies when either:
\begin{itemize}
\item[(1)] The GC system is strongly affected by that of a neighbouring galaxy. For example, the GC system of NGC4486A is difficult to separate from that of M87.
\item[(2)] The fitted parameters were unreliable due to a small numbers of GCs. Specifically, results were rejected when the error on the fitted effective radius exceeded 100\%.
\end{itemize}
Of our parent sample of 139 galaxies, we find 118 GC systems to have well measured GC density profiles with reliable model fits.

\subsection{Analysis of Globular Cluster Colors}
\label{data:colors}

Up to this point, the analysis has been carried out for the full GC system for each galaxy: i.e., with no restriction on GC colors. However, there is considerable interest in understanding how the size and structure of GC systems depends on both age and metallicity, so the fitting process was repeated after dividing each GC system into blue and red sub-components on the basis of their $(g'-i')_0$ colors.

To divide the sample, we fitted each GC color distribution using a Gaussian Mixture Modelling ({\tt GMM}, \citealp{Mur10}) code that identifies the dividing point in color between the blue and red GC systems.  If the $D$ value returned by GMM (a statistic that measures the separation of the means relative to their widths) was found to be smaller than 2.0, including errors, then we considered the GC system to be unimodal based on the manual of the GMM code. A more detailed study of the GC color distributions will be presented in future paper; for this analysis, we used these GC subsamples to explore of the sizes of the chemically distinct GC systems, fitting the separate radial number density profiles with S\'ersic profiles in the same way as for the full GC systems. Altogether we find 51 galaxies have statistically-significant bimodal GC color distributions.

Although separating the blue and red GC subsystems is relatively straightforward in high-mass galaxies, it can be problematic in intermediate- and low-mass systems --- a regime in which some galaxies have unimodal GC color distributions and, even for bimodal systems, the separation between the GC peaks can become subtle \citep{Pen06}. Moreover, the {\it interpretation} of these GC colors becomes problematic across a wide range in galaxy mass. In high mass galaxies, the often-seen blue and red GC sub-populations are respectively interpreted to be accreted and {\it in situ} populations of GCs. However, in dwarf galaxies, their often unimodal populations are almost always blue in color because of their low metallicity \citep[e.g.,][]{Lar22}. While this provides a link between the accreted population of GCs in more massive galaxies and their likely low-mass, metal-poor progenitors \citep[e.g.,][]{Cot98}, it obscures the origins of the GCs in dwarf galaxies themselves. Are the blue GCs in low-mass galaxies accreted as well? This is unlikely given the merger rate for dwarf galaxies, and the inefficiency of star cluster formation in the halos that accrete onto them. Given the similarity in color between dwarfs and their GCs, it is likely that the star formation episodes that formed the GCs also formed a significant fraction of the galaxy, and that they are thus associated.

For this reason, we introduce a new parameter to describe the color of GC sub-populations, $\Delta_{gi}$, that measures the difference between the color of the host galaxy and that of a (sub-)population of GCs. In the case of bimodal GC color distributions, a galaxy can have two $\Delta_{gi}$ values, one for each sub-population peak color, and for unimodal GC populations, it is simply the difference between the galaxy color and the median GC color. 

Figure~\ref{fig:colors} illustrates the definition of this parameter for two program galaxies. For galaxies with bimodal GC color distributions (as is the case for NGC4621, shown in the left panel), we denote the blue and red GC peak colors by $B$ and $C$. Relative to the underlying galaxy, which has a color\footnote{The colors of galaxies are mean colors within galaxy effective radii.} $A$, the GC sub-populations have values of $\Delta_{gi,BGC}=(B-A)$ for the blue GCs, and $\Delta_{gi,RGC}=(C-A)$ for the red GCs.
For galaxies with unimodal GC color distributions (such as NGC4612, shown on the right), we define a single parameter, $\Delta_{gi} = (B-A)$, as the difference between the median GC and galaxy color. We will return to the use of this parameter when discuss the color dependence of GC scaling relations. 
 
%%%%%%%%%%%%%%%%%%%%%%%%%%%%%%%%%%%%%%%%%%%%%%%%%%%%%%%%%%%%%%%%%%%%%%%%%%%%%%%%%%%%%%
%%%%%%%%%%%%%%%%%%%%%%%%%%%%%%%%%%%%%%%%%%%%%%%%%%%%%%%%%%%%%%%%%%%%%%%%%%%%%%%%%%%%%%
%%%
%%% RESULTS (Section 3) - Table 2, Figures 10-13
%%%
%%%%%%%%%%%%%%%%%%%%%%%%%%%%%%%%%%%%%%%%%%%%%%%%%%%%%%%%%%%%%%%%%%%%%%%%%%%%%%%%%%%%%%
%%%%%%%%%%%%%%%%%%%%%%%%%%%%%%%%%%%%%%%%%%%%%%%%%%%%%%%%%%%%%%%%%%%%%%%%%%%%%%%%%%%%%%

\section{Results}
\label{results}

Our program galaxies span a factor of $\sim$2000 in stellar mass and, as we show below, an equivalently wide range in GC system size, with $R_{e,{\rm GC}}\sim1$~kpc for dwarfs and $\sim100$~kpc for the brightest giants. In general, the blue GC systems have effective radii similar to those for the full GC systems while the largest red GC systems have effective radii of $\sim 30$~kpc --- much smaller than the blue or composite GC systems. Although the scatter is large, the measured S\'ersic indices mostly fall in the range $1 \lesssim n \lesssim 4$ with a median value of $n\sim2$ (albeit with fairly large measurement errors in most cases). The GC systems of some massive galaxies have large values of $n\sim4$, which is comparable to the galaxies themselves \citep{Fer06}. Errors on the measured effective radii are generally much smaller, with uncertainties of $\sim25\%$ being typical. $R_{e,{\rm gc}}$ will thus be our primary tracer of the spatial extent of the GC systems. In this section, we examine how GC system effective radius scales with the properties of the host galaxy and dark matter halo.

%%%%%%%%%%%%%%%%%%%%%%%%%%%%%%%%%%%%%%%%%%%%%%%%%%%%%%%%
\begin{figure}
\epsscale{1.15}
\plotone{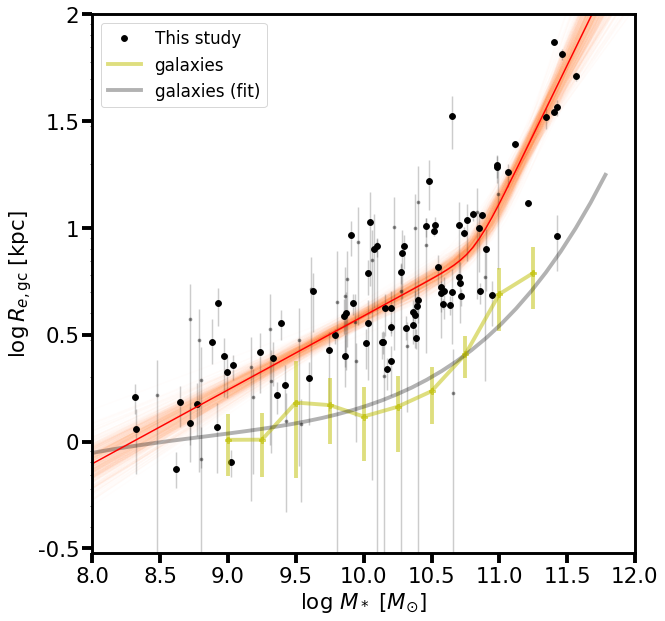}
\caption{The effective radii of GC systems plotted against the stellar masses of their host galaxies.  Filled black circles show GC systems from this study; well-fitted systems (error of $R_{e,gc} < 50\%$) are shown with large heavy symbols.  Stellar masses for Virgo galaxies are taken from Roediger et al. (2023 in prep.). Stellar masses for MATLAS galaxies are taken from ATLAS$^{\rm 3D}$, but adjusted to match the NGVS results computed using the subsample of galaxies in common to ATLAS$^{\rm 3D}$ and NGVS (${\rm log}M_{*,NGVS} = {\rm log}M_{*,{\rm ATLAS}^{\rm 3D}} - 0.11$). The heavy orange line shows the best-fit broken power-law function to the well-fitted systems; light orange lines show 500 random MCMC samples based on this fit.  
The relation between effective radius and stellar mass for our sample galaxies are shown by the green symbols (binned data) and gray solid curve (model, C{\^o}t{\'e} et al. 2023, in prep.), respectively.
\label{fig:sizemstar}}
\end{figure}
%%%%%%%%%%%%%%%%%%%%%%%%%%%%%%%%%%%%%%%%%%%%%%%%%%%%%%%%

\subsection{Scaling Relations: I. Galaxy Parameters}
\label{results:scale_galaxy}
\subsubsection{$R_{e,gc}$ vs.\ Galaxy Stellar Mass}

Since stellar mass is a fundamental parameter that dictates many other galaxy properties, we begin our analysis by considering the the dependency of GC system effective radius on the stellar mass of the host galaxy. The measurements that form the basis of our analysis are recorded in Table 2 of Lim et al. (2024 in preparation). 

\begin{deluxetable}{cccc}
\tablecaption{GC System Size versus Stellar Mass Relation
\label{tab:fittedpars}}
\tablehead{
\colhead{ } & \colhead{All GCs} & \colhead{Blue GCs} & \colhead{Red GCs}
}
\startdata
$M_p$ [$10^{10} M_{\odot}]$ & $6.5^{+1.9}_{-1.5} $ & $24^{+19}_{-12}$ & $4.5^{+1.4}_{-1.0}$ \\
$R_p$ [kpc] &$8.3^{+1.6}_{-1.2}$  & $32^{+13}_{-13}$ & $4.6^{+0.8}_{-0.6}$ \\
$\alpha$ & $0.34^{+0.04}_{-0.04}$  & $0.56^{+0.04}_{-0.05}$ & $0.16^{+0.05}_{-0.05}$ \\
$\beta$ & $1.30^{+0.22}_{-0.17} $ & $1.14^{+0.58}_{-0.55}$ & $0.84^{+0.15}_{-0.13}$ \\
\enddata
\end{deluxetable}

Figure~\ref{fig:sizemstar} shows the effective radius of the full GC system plotted against the stellar mass of the host galaxy. 
Results from this study are plotted as black symbols.
For comparison, the gray curve shows the best-fit polynomial for galaxies in the Virgo core region from C{\^o}t{\'e} et al. (2023, in prep.). This relation exhibits the familiar steepening of the galaxy size-mass relation above $\sim$10$^{10.5}M_{\odot}$. We can see from Figure~\ref{fig:sizemstar} that this behavior is mirrored by the GC systems of galaxies in this same environment (Virgo), with an unmistakable ``break" in the size-mass relation at intermediate masses. 

%%%%%%%%%%%%%%%%%%%%%%%%%%%%%%%%%%%%%%%%%%%%%%%%%%%%%%%%
\begin{figure}
\epsscale{1.15}
\plotone{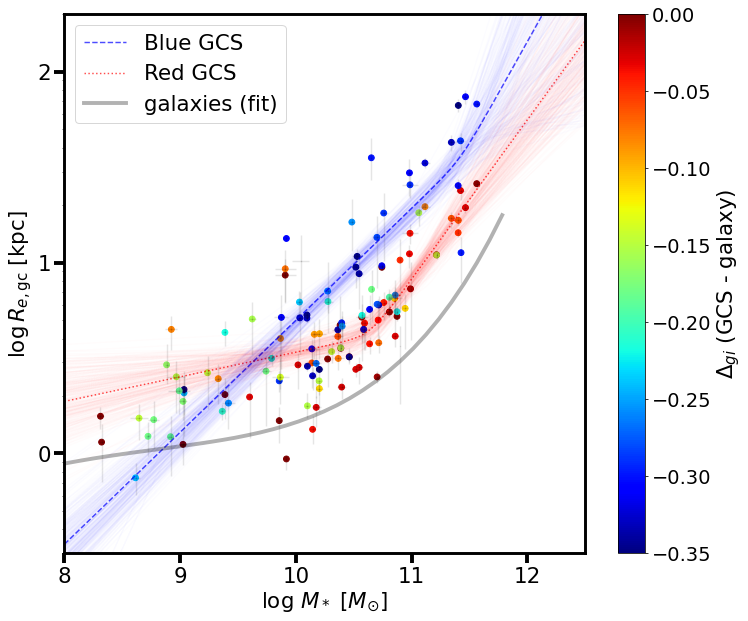}
\caption{The effective radii of blue and red GC subsystems plotted against the stellar masses of their host galaxies. The data points show the effective radii for blue and red GC systems, or for the total GC system of galaxies with unimodal GC color histograms. Data points are color-coded by the difference in color between the GC system and host galaxy as shown by the color bar at the right. 
GC systems with well fitted effective radii are only shown.
The blue dashed curve shows the best-fit broken power-law relation for blue GCs. The red dotted curve shows the best-fit result for red GCs. In both cases, the light curves show  500 random MCMC samples of based on these best-fit results. The gray curve shows the relation between effective radius and stellar mass for NGVS galaxies (C{\^o}t{\'e} et al. 2023, in prep.).
\label{fig:sizemstarcolor}}
\end{figure}
%%%%%%%%%%%%%%%%%%%%%%%%%%%%%%%%%%%%%%%%%%%%%%%%%%%%%%%%

In this log-log representation, the trend is clearly non-linear; to capture this behaviour, we fitted the data using a broken power-law function \citep{Mow19} that has been used previously to analyze the size-mass scaling relation of galaxies:
\begin{equation}
\label{dblpower}
R_{e,{\rm gc}} = R_{p} \left(\frac{M_*}{M_p}\right)^\alpha \left[ \frac{1}{2} \left\{ 1 + \left( \frac{M_*}{M_p} \right)^\delta \right\} \right]^{(\beta - \alpha)/\delta}
\end{equation}
Here $R_p$ is the ``pivot" radius of the GC system which marks the change in slope, $M_p$ is the pivot stellar mass, $\alpha$ is the slope at the low-mass end, $\beta$ is the slope at the high-mass end, and $\delta$ is a smoothing factor. 
After some experimentation, we opted to fix the smoothing parameter to $\delta = 6$ rather than allowing it to vary as a free-parameter (see also \citealt{Mow19}).

%%%%%%%%%%%%%%%%%%%%%%%%%%%%%%%%%%%%%%%%%%%%%%%%%%%%%%%%
\begin{figure*}[htbp]
\centering
\includegraphics[width=0.99\textwidth]{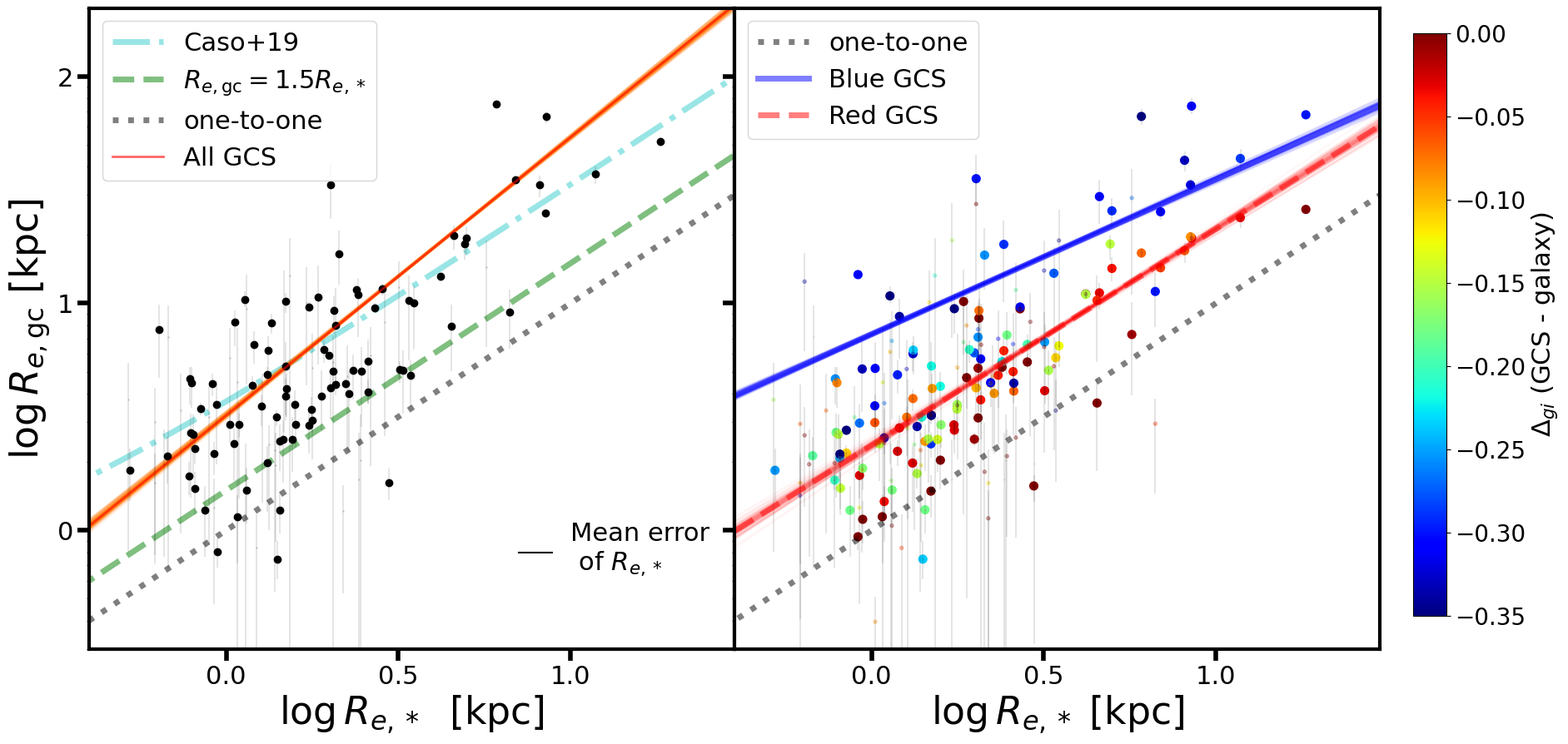}
\caption{(Left panel) The effective radii of GC systems plotted against the effective radii of their host galaxies. Measurements for the full GC system are shown as black symbols. The heavy orange line shows the best-fit linear relation in log-log space. The light orange lines represent 100 random resampling results based on this fitted relation. The gray dashed line shows one-to-one relation. The green dashed line shows the relation $R_{e,gc} = 1.5 R_{e,*}$ which has sometimes adopted in studies of the GC systems belonging to ultra-diffuse galaxies \citep{van16,Lim18}. The cyan dashed line shows the relation from \citet{Cas19}. The typical error bar of galaxy effective radii is noted at the bottom right. (Right panel) Same as in the previous panel, except the GC systems have been divided by color. Symbols colors are same as in Figure~\ref{fig:sizemstarcolor}. The blue dotted and red dashed curves show the best-fit linear relations in log-log space for blue and red GCs, respectively. The light blue and light red curves shown 100 random re-sampling results based on these fits.
\label{fig:sizesize}}
\end{figure*}
%%%%%%%%%%%%%%%%%%%%%%%%%%%%%%%%%%%%%%%%%%%%%%%%%%%%%%%%

Our fitting results are summarized in Table~\ref{tab:fittedpars}. 
The slope we measure at the high-mass end, $\beta = 1.30^{+0.11}_{-0.17} $, is consistent with values found in previous studies: $1.30\pm0.14$ \citep{Hud18} and $0.97\pm0.4$ \citep{For17}. The slope at the low-mass end --- a regime that is largely unexplored --- is significantly shallower than this, with $\alpha = 0.34^{+0.04}_{-0.04}$. We find the break in the size-mass relation to occur at a stellar mass of $M_p = 6.5^{+1.9}_{-1.5} \times 10^{10} M_{\odot}$. 
\citet{Cas19} reported a similar estimate for this transition  (i.e., $\sim 4\times10^{10}M_{\odot}$) although the functional form they adopted does not fit the data well at low and high masses (see \S\ref{sec:connectDM}). 

Figure~\ref{fig:sizemstarcolor} shows these same relations for the blue and red GC systems. In this plot, symbols for individual GC systems --- either their separate blue and red components when bimodal, or the total color distribution, if unimodal --- have been color-coded by the $\Delta_{gi}$ parameter. This parameter, which has been discussed in \S\ref{data:colors}, measures the color offset relative to the underlying galaxy. 
In this plot, we see that the size--mass relation separates into two branches at high masses, having distinct ``blue" and ``red" $\Delta_{gi}$ values. Below stellar masses of  $\sim 10^{10}M_{\odot}$, where unimodal or weakly bimodal GC systems are common, $\Delta_{gi}$ values are intermediate in color, with many ``green" values. In this regime, the GC systems appear to define a single, shallow relation.

Best-fit broken power-law parameters for the blue and red GC systems are recorded in  Table~\ref{tab:fittedpars}. 
These were estimated using a weighted fit, where each point was weighted by both its uncertainty {\it and} its $\Delta_{gi}$ parameter. Therefore, there is no strict cut that includes or excludes data points from each fit. For example, the power law fit for the blue GC systems includes all points, but gives greater weight to the ones that have bluer $\Delta_{gi}$. For galaxies with bimodal GC color distributions, this effectively gives the red modes nearly zero weight, whereas the unimodal GC color distributions are often given an intermediate weight.

As a rule, nearly all blue GC systems are larger in size than the red GC systems within the same host galaxy. 
The trends for the blue and red GC systems show differences from that of the composite GC systems.
Although the slope of blue GC systems at high-mass, and the break point of red GC systems, are consistent with those of the composite GC systems, the slope of low-mass blue GC systems, the break point of blue GC systems, and both slopes of the red GC systems are different with those of composite GC systems. 
Interestingly, the slope at low-mass for composite GC systems is consistent with a mean value of the slopes at low-mass for blue and red GC systems. 
The above results suggest that the break observed in composite GC systems is not only driven by a transition to having more blue GCs at higher mass, but also by an increase in size of red GC systems (mirroring the stars). We will discuss the implications of this finding in \S\ref{discussion}.

%%%%%%%%%%%%%%%%%%%%%%%%%%%%%%%%%%%%%%%%%%%%%%%%%%%%%%%%
\begin{figure*}[htbp]
\centering
\includegraphics[width=0.99\textwidth]{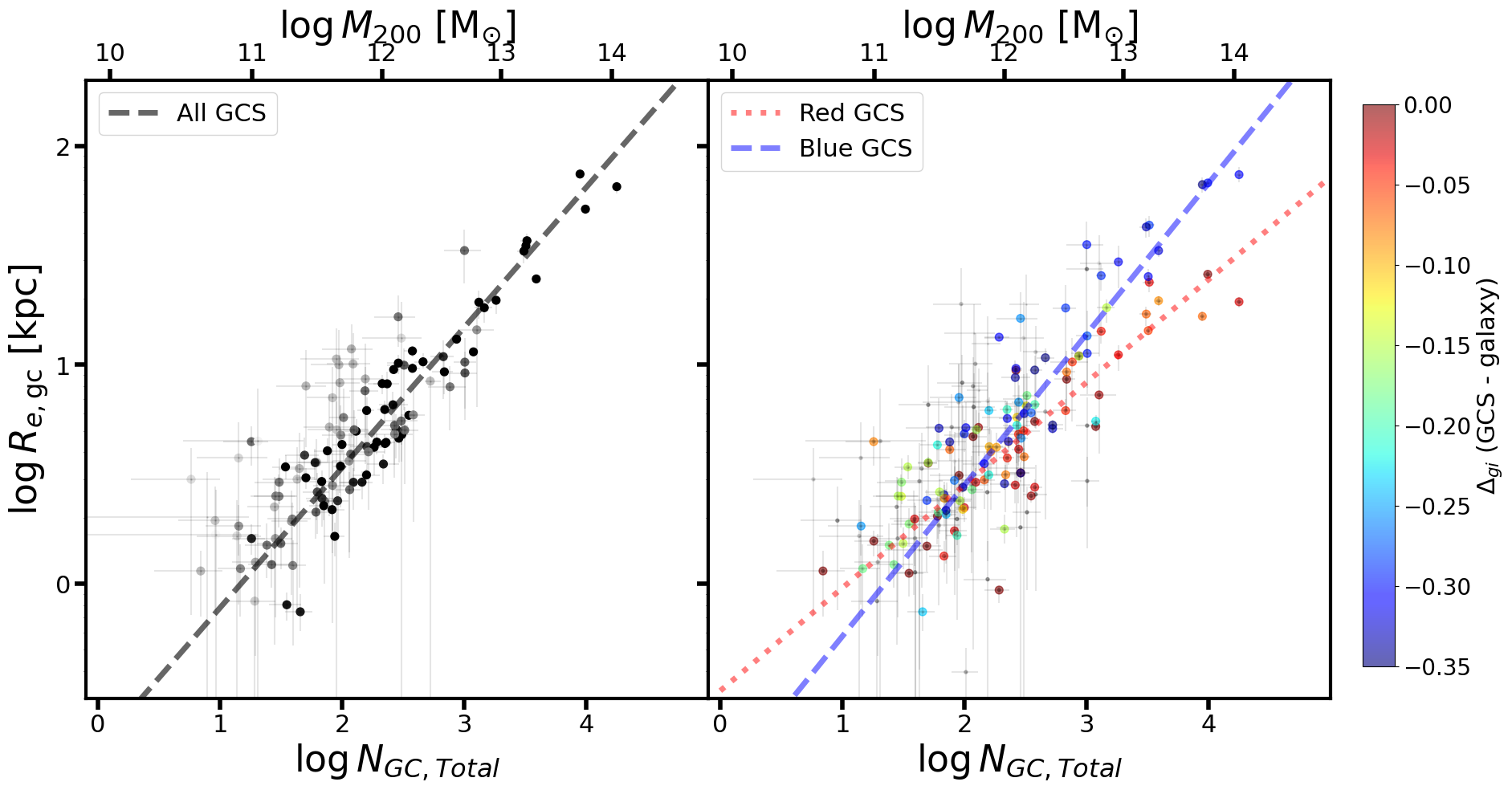}
\caption{(Left panel) The effective radii of GC systems plotted against the total number of GCs. Measurements for the full GC systems are shown as black symbols; the large, heavy symbols show those galaxies with well-fitted GC density profiles. The best-fit relation is shown by the black line. The upper x-axis shows the dark matter halo mass ($M_{200}$) corresponding to the computed number of GCs in each galaxy. (Right panel) Same as in the previous panel, except symbols have been color-coded by the color difference between GC systems and their host galaxy, as indicated by the color bar on the right. The blue and red lines show the best-fit fitted linear functions for the blue and red GC subsystems. See \S\ref{data:fitting} for details on the separation of blue and red GCs.
\label{fig:numsize}}
\end{figure*}
%%%%%%%%%%%%%%%%%%%%%%%%%%%%%%%%%%%%%%%%%%%%%%%%%%%%%%%%

\subsubsection{$R_{e,gc}$ vs.\ Galaxy Size}

We now turn our attention to the variation in the relative sizes of the GC systems and underlying galaxies. Figure~\ref{fig:sizesize} shows a comparison between the effective radii of the GC systems, $R_{e,{\rm gc}}$, and those of their host galaxies, $R_{e,*}$. In the left panel, the black symbols show results based on the size of the full GC system; on the right, we show results for the blue and red GC systems separately, with the symbols color-coded as in the right panel of Figure~\ref{fig:sizemstarcolor}. The orange line in the left panel, and the blue and red lines in the right panel, show the respective best-fit linear relations in log-log space:

\footnotesize
\begin{eqnarray}
\log{R_{e,{\rm gc}}} & = & (1.22\pm0.01)\log R_{e,*} + (0.51\pm0.01)~~~{\rm all}\\
                              & = &  (0.68\pm0.02)\log R_{e,*} + (0.87\pm0.01)~~~{\rm blue}\\
                              & = &  (0.95\pm0.02)\log R_{e,*} + (0.38\pm0.01)~~~{\rm red}
\end{eqnarray}
\normalsize
The cyan line in the left panel shows the relation of \cite{Cas19}\footnote{Note that \cite{Cas19} fitted their data in linear units, rather than log-log space as we have done.}.
Comparing our measurements to the one-to-one relation (which is shown as the dotted line in each panel), the GC systems are observed to be significantly more extended the the galaxies themselves --- conclusion that holds whether one considers the composite GC system, or separate the blue and red subcomponents. We also note that the trend for red GCs appears to show less scatter than that for the blue GCs ($rms$ = 0.03 vs. 0.06). A comparison of Figure~\ref{fig:sizemstar} and \ref{fig:sizesize} shows the correlation between GC size and stellar mass to be significantly tighter than the correlation with host galaxy size. Of course, much of the scatter in Figure~\ref{fig:sizesize} may be attributable to measurement errors in $R_{e,*}$, which can be significant (see, e.g., \citealt{Che10}).

Finally, we point out that the green dashed line in Figure~\ref{fig:sizesize} shows the relation $R_{e,{\rm gc}} = 1.5R_{e,{*}}$, which has sometimes been adopted as an estimator of GC system size, most notably in the study of ultra-diffuse galaxies (e.g., \citealt{van16}). As this figure shows, this relation significantly underestimates the size of the GC system. If we fix the slope of fitted relation at unity in log-log spaces, we obtain a best-fit scaling relation of $R_{e,{\rm gc}} = (4.5\pm0.1)R_{e,{*}}$ for all GCs. If we consider the blue and red GCs separately, then we find factors of $7.9\pm0.1$ and $2.2\pm0.1$, respectively.

Here we show that even the red GC population---i.e., the ones with colors most closely matched to those of the host galaxy population and are most likely to have formed {\it in situ}---have an extent that is twice as large as the underlying stars (on average), and that this is true over a wide range of mass. Given that intense star formation episodes in which GCs are most likely to form are often centrally concentrated, this result requires that even GCs that form {\it in situ} encounter evolutionary mechanisms (whether formation or destruction) that bias their spatial distribution toward a galaxy's outer regions. We note that we all of our measurements are for roughly the brighter half of the GC luminosity function, and that if there are mass-dependent processes (such as dynamical friction), then deeper studies might reveal a different radial distribution for fainter GCs.

\subsection{Scaling Relations: II. Dark Matter Parameters}
\label{results:scale_dm}

Having established in the previous section that the sizes of GC systems --- including those of their blue and red subcomponents --- are correlated with the stellar masses of their host galaxies, we now turn our attention to possible connections with the total numbers of GCs and, by extension, the masses of their host dark matter halos. Our discussion of these parameters is linked by the assumption that the total number of GCs belonging to a galaxy is directly proportional to the total mass of the galaxy, which is in turn dominated by the dark matter halo. This possibility, which dates back to \cite{Bla97} and \cite{McL99}, has gained momentum in recent years, with several studies providing support for the claim (e.g., \citealt{Bla99,Hud14,Har15,For16,Har17}).

Figure~\ref{fig:numsize} shows the dependence of the GC system effective radius on the total number of GCs in each galaxy, $N_{GC, Total}$, using data from Table 2 of Lim et al. (2023 in prep.). 
We emphasize that both of these parameters are measured directly from our GC density profile fits: i.e., the effective radius for each GC system is measured as a free parameter, while the total number of GCs is calculated by numerically integrating the background-corrected density profile (Equation~1). Following this integration, we correct the GC numbers for our magnitude limit in each galaxy using the dependence of the GC luminosity on host galaxy luminosity \citep{Jor07,Vil10}. Figure~\ref{fig:numsize} presents the result of this exercise in the same way as Figure~\ref{fig:sizesize}: i.e.,  the left panel shows results for the full GC system while results for the individual blue and red GC systems are shown on the right.

It is apparent that galaxies show a tight correlation between the effective radii of their GC systems and the total number of GCs. This is not surprising given that the two parameters are related to each other via the S\'ersic law used in our analysis (although the concentration of the GC system, as measured by S\'ersic index, and the density at the effective radius also contribute to the total system richness). However, we stress that, in Figure~\ref{fig:numsize}, the two-dimensional error ellipses for individual points have randomly distributed ellipticities and position angles. This indicates that correlated errors are certainly not responsible for the observed trend, which spans two full decades in effective radius and more than three decades in $N_{GC, Total}$.

We fitted the relations in Figure~\ref{fig:numsize} with a linear function in log-log space:
\footnotesize
\begin{eqnarray}
\log R_{e,{\rm gc}} & = &  (0.65\pm0.02)\log N_{GC, Total} - (0.77\pm0.06)~~~{\rm all}\\
                              & = &  (0.69\pm0.042\log N_{GC, Total} - (0.92\pm0.06)~~~{\rm blue}\\
                              & = &  (0.47\pm0.03)\log N_{GC, Total} - (0.48\pm0.08)~~~{\rm red}
\end{eqnarray}
\normalsize
The tightness of this relation was previously noted by \citet{Cas19}. However, they found the trend to flatten at low galaxy masses (corresponding to $N_{GC, Total}\lesssim 100$) and opted to fit the relation using a second order polynomial. In this study, we find that a single linear relation (shown as the dashed line in the left panel) provides an excellent representation of the data from $N_{GC, Total}\sim 10$ to $\sim 10000$.

The right panel of this figure shows this same relation but now replaced with the effective radii of the blue and red GC subcomponents. At the high-mass end, there is a clear separation between the blue and red GC systems. However, at low masses, the separate branches disappear entirely and only a single relation is apparent below $N_{GC, Total}\sim 300$, with the best-fit relations formally crossing at $N_{GC, Total} \sim100$. In other words, the blue and red GC systems belonging to these low-mass galaxies no longer show distinct spatial distributions. We will return to this issue in \S\ref{discussion}, since it may support the view that blue GCs in these low-mass hosts represent bonafide ``{\it in-situ}" populations from which the GC systems of higher-mass galaxies have been assembled.

Finally, the horizontal axes along the top of both panels in Figure~\ref{fig:numsize} indicate the halo mass, $M_{\rm dm}$, corresponding to the total number of GCs measured from our density profile fits and plotted along the lower horizontal axes. To convert from $N_{GC, Total}$ to $M_{\rm dm}$, we used the relation of \cite{Har17}
\begin{eqnarray}
\label{eq:eta}
\eta_M \equiv \langle{m_{\rm gc}}\rangle{N_{GC, Total}} / M_{\rm dm} & = & (2.9\pm0.2)\times10^{-5}
\end{eqnarray}
and adopted the same average GC mass of $\langle{m_{\rm gc}}\rangle = 2.8\times10^5M_{\odot}$ used in that study. \citet{Har17} established this relation based on the halo mass derived from K-band galaxy luminosity, calibrated through gravitational lensing, and assuming a mean mass-to-light ratio of GCs with $\langle M(GCS)/L_V \rangle \sim1.3$. They used the total number of GCs from compiling literature data. If we choose to fit the log-log relations shown in Figure~\ref{fig:numsize} with $\log{M_{\rm gc}}$ as the independent variable, we find

\footnotesize
\begin{eqnarray}
\log R_{e,{\rm gc}} & = &  (0.65\pm0.02)\log M_{\rm dm} - (7.20\pm0.24)~~~{\rm all}\\
                              & = &  (0.69\pm0.02)\log M_{\rm dm} - (7.78\pm0.27)~~~{\rm blue}\\
                              & = &  (0.47\pm0.03)\log M_{\rm dm} - (5.15\pm0.33)~~~{\rm red}
\end{eqnarray}
\normalsize

With halo masses in hand, we now turn our attention to an exploration of how the GC system properties correlate with various halo parameters, and what those correlations imply for the formation of galaxies and their GC systems.

%%%%%%%%%%%%%%%%%%%%%%%%%%%%%%%%%%%%%%%%%%%%%%%%%%%%%%%%%%%%%%%%%%%%%%%%%%%%%%%%%%%%%%
%%%%%%%%%%%%%%%%%%%%%%%%%%%%%%%%%%%%%%%%%%%%%%%%%%%%%%%%%%%%%%%%%%%%%%%%%%%%%%%%%%%%%%
%%%
%%% DISCUSSION (Section 4) - Figures 14-18
%%%
%%%%%%%%%%%%%%%%%%%%%%%%%%%%%%%%%%%%%%%%%%%%%%%%%%%%%%%%%%%%%%%%%%%%%%%%%%%%%%%%%%%%%%
%%%%%%%%%%%%%%%%%%%%%%%%%%%%%%%%%%%%%%%%%%%%%%%%%%%%%%%%%%%%%%%%%%%%%%%%%%%%%%%%%%%%%%

\section{Discussion}
\label{discussion}

\subsection{A Check on Halo Masses}
\label{section:check}

%%%%%%%%%%%%%%%%%%%%%%%%%%%%%%%%%%%%%%%%%%%%%%%%%%%%%%%%
\begin{figure}
\epsscale{1.15}
\plotone{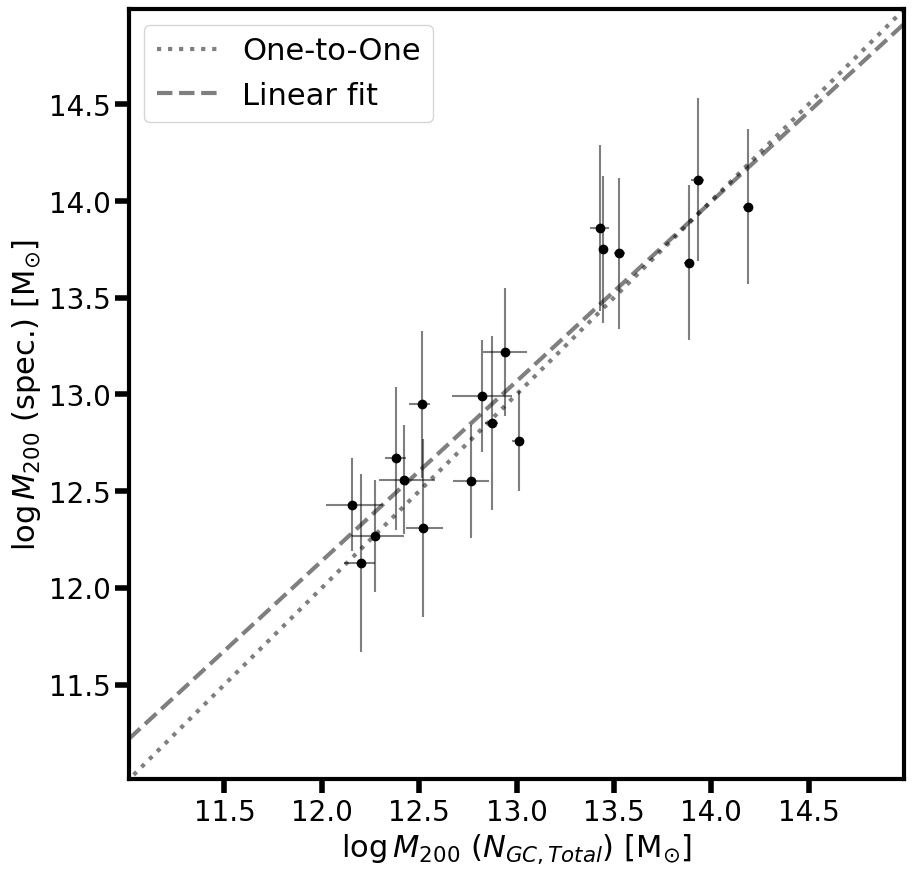}
\caption{Comparison of halo masses for the subset of our sample galaxies with dynamical halo mass measurements in the literature. The abscissa shows halo masses inferred from the observed number of GCs,  $N_{GC, Total}$, as described in \S\ref{results:scale_dm}. The ordinate shows dynamically measured halo masses for a sample of 18 galaxies from the SLUGGS survey \citep{Ala17}. The dotted line shows the one-to-one relation, while the dashed line shows the best-fit linear relation.
\label{fig:dmcomparison}}
\end{figure}
%%%%%%%%%%%%%%%%%%%%%%%%%%%%%%%%%%%%%%%%%%%%%%%%%%%%%%%%

In this section, we use our measurements to examine the connection between the GC systems and dark matter halos associated with our program galaxies. Of course, we do not probe the halos $directly$, so this discussion, like many before it, hinges on the {\it ansatz} that the total number of GCs belonging to a galaxy scales in proportion to the halo mass. Before proceeding with the discussion, we therefore pause to perform a simple check on this assumption.

Because our sample includes a number of bright, nearby, well studied galaxies, we can test the key assumption from \S\ref{results:scale_dm} that the observed number of GCs, $N_{GC, Total}$ is linearly related to the halo mass (i.e., Equation~\ref{eq:eta}). Figure~\ref{fig:dmcomparison} shows a comparison, for 18 of our program galaxies, between the halo mass derived in this way (plotted on the abscissa) and the dynamically measured halo mass (plotted on the ordinate).\footnote{The NGC identifications of these 18 galaxies are: 4472, 4649, 4374, 4365, 4473, 4459, 4564, 4474, 821, 1023, 2768, 3607, 3608, 4278, 5846, 5866 and 7457.} Dynamical masses are taken from the SLUGGS survey \citep{Bro14} and are based on GC radial velocity measurements.  These masses were derived by using the tracer mass estimator of \cite{Wat10} to derive homogeneous masses within 5$R_e$. These masses were then corrected to $M_{\rm 200}$ (i.e., Table~3 of \citealt{Ala17}).

The dotted line in Figure~\ref{fig:dmcomparison} shows the one-to-one relation, which should reflect the trend if our assumption is correct that  $N_{GC, Total}$ is an accurate tracer of halo mass. This does indeed appear to be the case, with unweighted best-fit relation showing good agreement with the expected relation (with an $rms$ scatter of 0.23 dex). Although this comparison is limited in that the plotted galaxies span only $\sim$ two orders of magnitude in halo mass, as opposed to the full sample which nearly four orders of magnitude (see Figure~\ref{fig:numsize}), we conclude that the available evidence suggests Equation~\ref{eq:eta} is a faithful predictor of halo mass for our program galaxies.

\subsection{Connection to Dark Matter Halo Parameters}
\label{sec:connectDM}
%%%%%%%%%%%%%%%%%%%%%%%%%%%%%%%%%%%%%%%%%%%%%%%%%%%%%%%%
\begin{figure*}[htbp]
\centering
\includegraphics[width=0.99\textwidth]{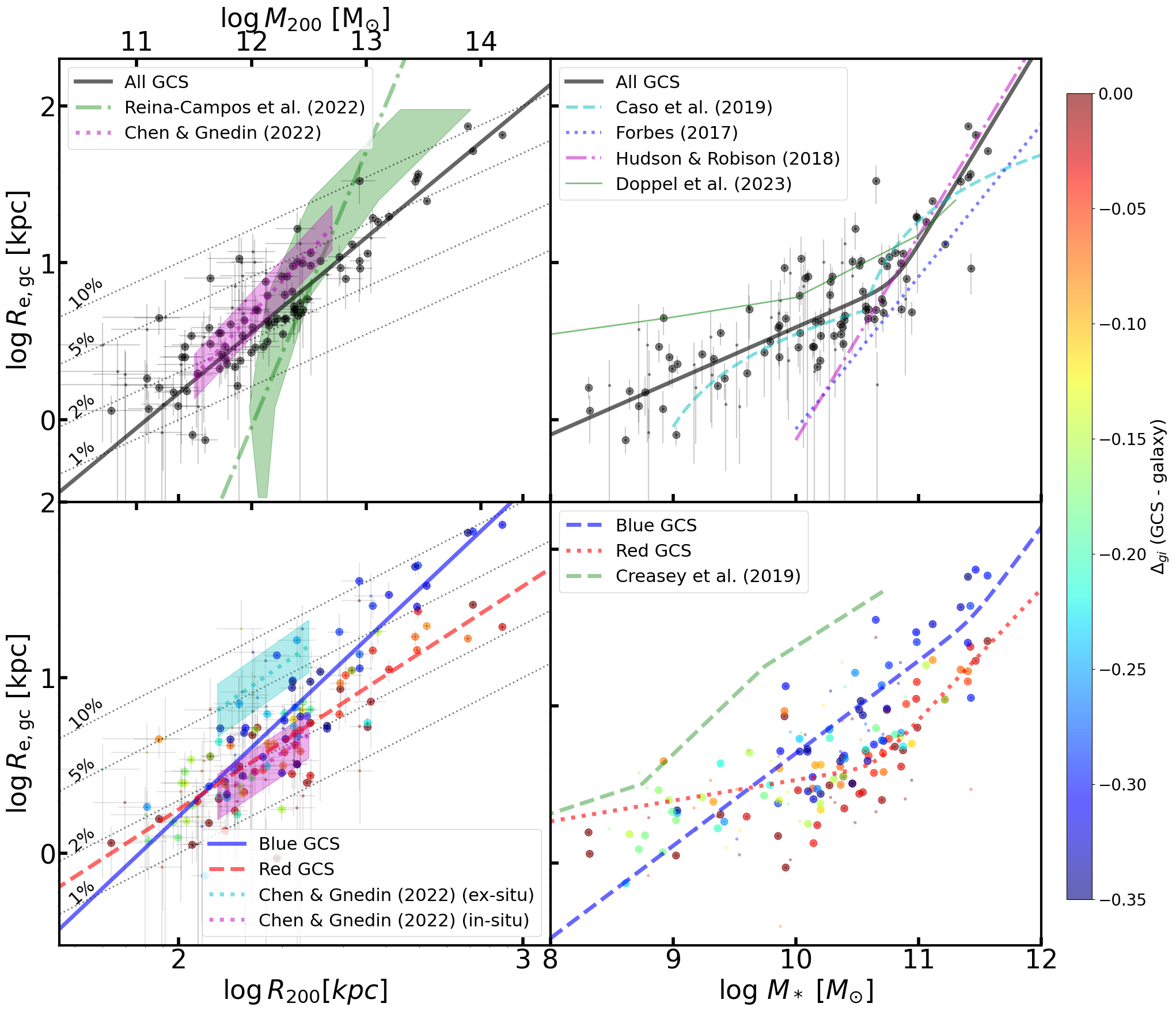}
\caption{Comparison with various models and other observations. All y-axes represent the effective radius of the GC system, $R_{e,{\rm gc}}$. Panels in the left column show relations between the virial radius, $R_{\rm 200}$, and $R_{e,{\rm gc}}$, while panels in the right column show relations between host galaxy stellar mass, $M_*$, and $R_{e,{\rm gc}}$. We calculated $R_{\rm 200}$ directly from the halo mass (which itself was based on $N_{gc,Total}$), so corresponding halo masses are noted on the top for reference. All panels' solid black, blue, and dashed red lines show the best-fit relations for the total, blue, and red GC systems. The grey dotted lines in each of the left column panels are drawn at fixed fractions of ${\rm log}R_{200}$ (i.e., 1\%, 2\%, 5\%, 10\%, 20\% or 50\%). Data points in the bottom panels are color-coded, as shown in the right color bar. (Top left) The dot-dash green line shows a relation based on the simulated GC systems from E-MOSAICS \citep{Rei21} while the shaded green band in each panel corresponds to the median and 25--75th percentiles in each parameter from the simulations. The magenta dotted line and shaded regions shows a relation from simulated GC systems based on the Illustris TNG50-1 \citep{Che22}. (Bottom left) Cyan and magenta dotted lines with shaded regions represent relations for {\it in situ} and {\it ex situ} simulated GC systems from \citet{Che22}, respectively. (Top right) Dashed cyan \citep{Cas19}, dotted blue \citep{For17}, and dashed-dot magenta lines \citep{Hud18} represent previous observational results, whereas the light green solid line shows a simulation result from \citet{Dop23} based on Illustris TNG50. (Right bottom) The green dashed line shows the relation from the simulation by \citet{Cre19}.  
\label{fig:models}}
\end{figure*}
%%%%%%%%%%%%%%%%%%%%%%%%%%%%%%%%%%%%%%%%%%%%%%%%%%%%%%%%

%%%%%%%%%%%%%%%%%%%%%%%%%%%%%%%%%%%%%%%%%%%%%%%%%%%%%%%%
\begin{figure*}
\epsscale{1.15}
\plotone{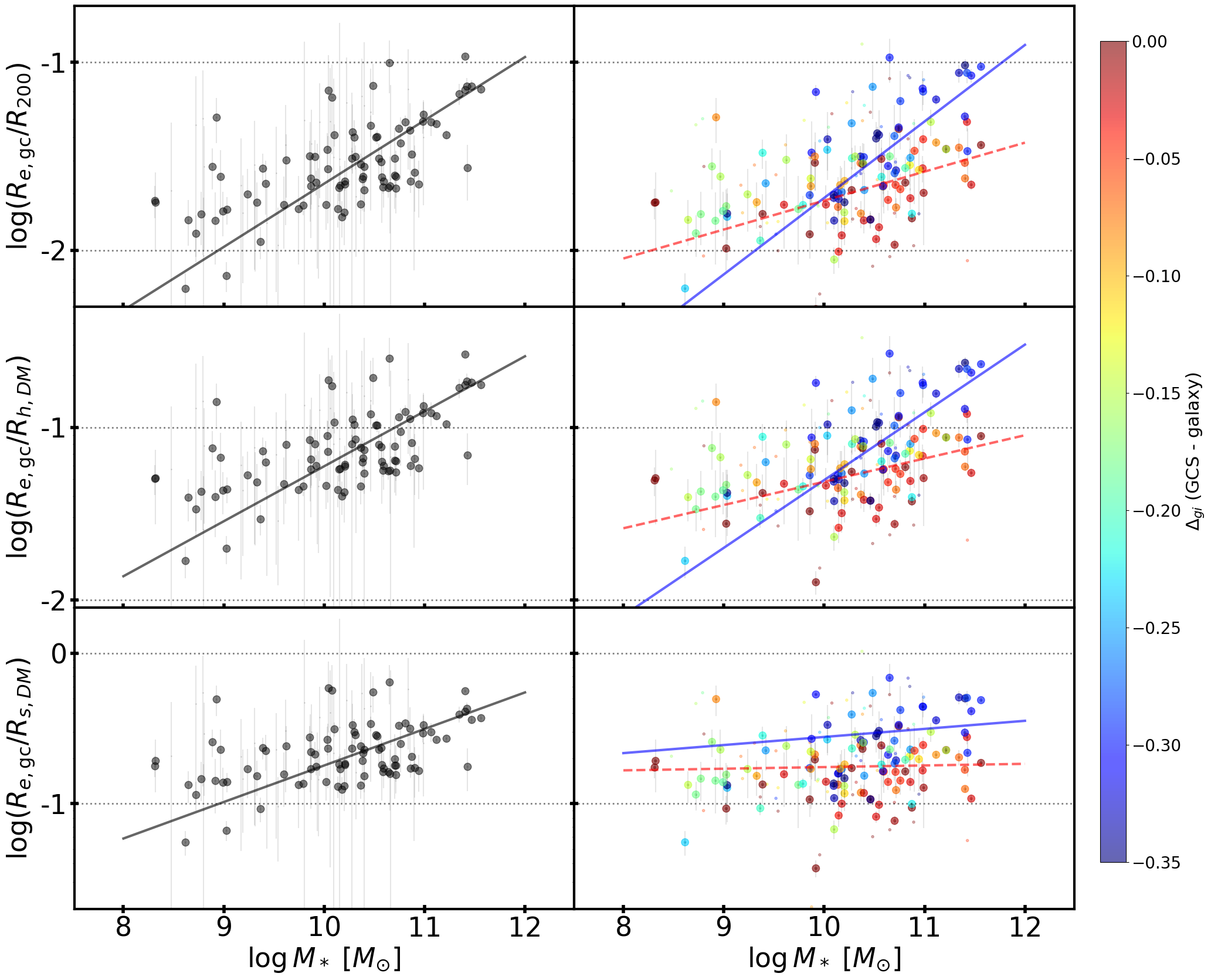}
\caption{Variation in globular cluster system size as a function of host galaxy stellar mass. The panels on the left show results for the full globular cluster systems (black symbols). Results for the separate blue and red globular clusters are shown in the right panels. For both columns, the globular cluster effective radius is normalized to three different measures of the size of the dark matter halo: i.e., the virial radius, $R_{\rm 200}$, the radius containing half the virial mass, $R_{\rm h,dm}$, and the NFW scale radius, $R_{\rm {S,dm}}$ (top to bottom, respectively).
\label{fig:fg_dm_mstar}}
\end{figure*}
%%%%%%%%%%%%%%%%%%%%%%%%%%%%%%%%%%%%%%%%%%%%%%%%%%%%%%%%

By fitting the GC density profiles for our program galaxies, we determine two fundamental parameters for each GC system: (1) the total number of GCs, $N_{GC, Total}$, which is used to infer the dark matter halo mass (\S\ref{results:scale_dm}); and (2) the effective radius, $R_{e,{\rm gc}}$, which allows us to compare the spatial extent of the GC system to that of the underlying halo. For this comparison, we use three different radial measures for the halo, which is assumed to have an NFW profile in all galaxies. 

First, we use our GC-based estimate for the mass of the halo, $M_{200}$, to calculate $R_{200}$ (e.g., Equation~7 of \citealt{Lok01})
\begin{eqnarray}
\label{eq:lm1}
M_{200} & = & {4 \over 3}{\pi}rR_{200}^3{v}{\rho{_c^0}}
\end{eqnarray}
with $\rho{_c^0} = 277.5~M_{\odot}~{\rm pc}^{-3}$, and virial overdensity, $v=200$. Our second measure for the spatial extent of the halo, the scale radius, $R_s$, then follows from $R_s = R_{200}/c_{200}$ with the concentration parameter given by Equation~9 of \cite{Dut14}:
\begin{eqnarray}
\label{eq:lm1}
\log c_{200} & = & 0.905 - 0.101\log [M_{200} / (10^{12}~h^{-1}~M_{\odot})]
\end{eqnarray}
Finally, we solve numerically Equation~8 of \cite{Lok01},
\begin{eqnarray}
\label{eq:lm1}
M(s)/M_{\rm v} & = & g(c) [ \ln(1+ cs) - {cs / (1+cs)} ],
\end{eqnarray}
where 
\begin{eqnarray}
\label{eq:lm1}
g(c) & = & {1 \over {\ln{(1 + c)}} - c/(1+c)}
\end{eqnarray}
to find the radius, $R_h$, containing half the halo mass.
This provides us with three radial scales for the assumed NFW profile: $R_s$, $R_h$ and $R_{200}$.

%%%%%%%%%%%%%%%%%%%%%%%%%%%%%%%%%%%%%%%%%%%%%%%%%%%%%%%%
\begin{figure*}
\epsscale{1.15}
\plotone{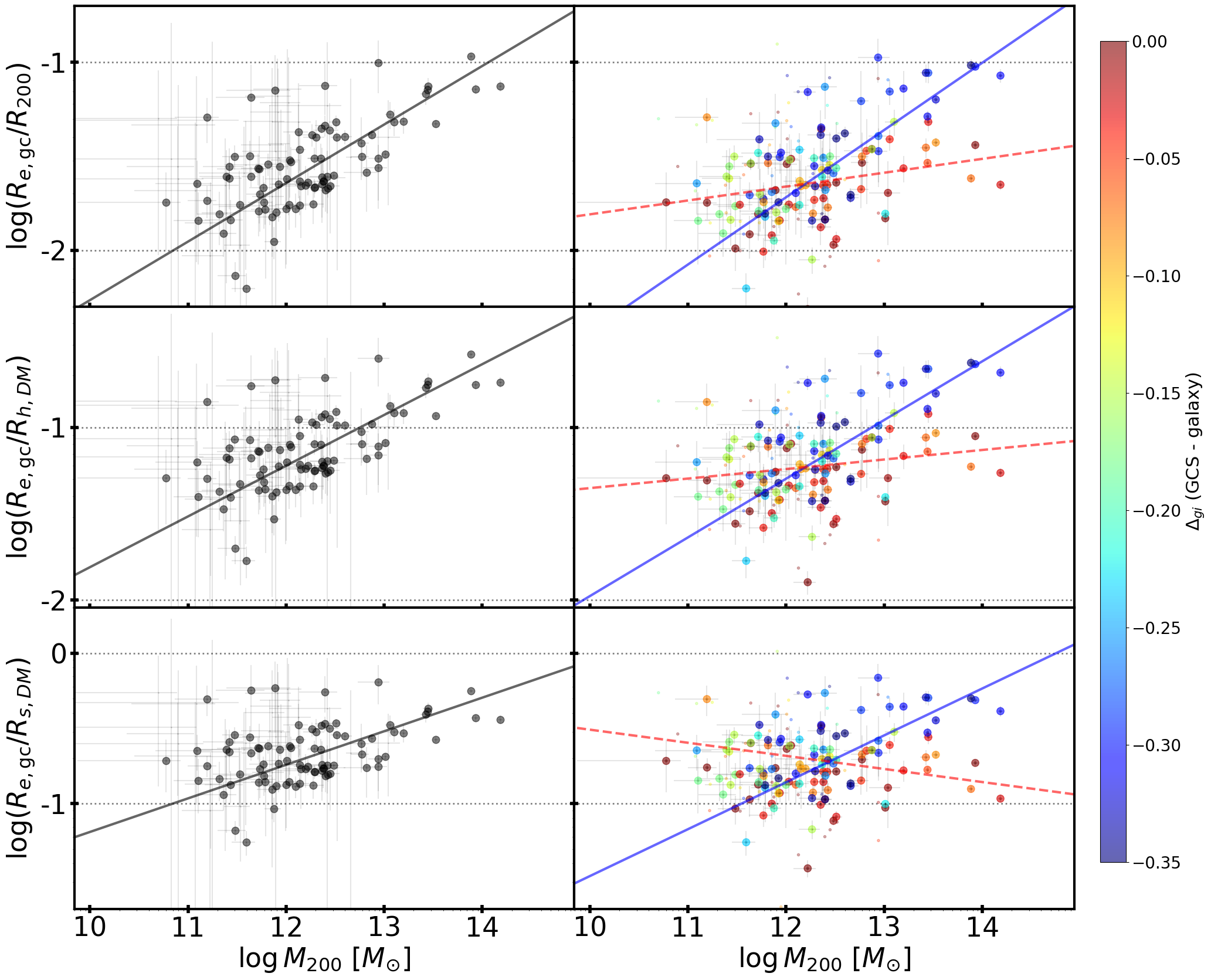}
\caption{Variation in globular cluster system size as a function of inferred halo mass. The panels on the left show results for the full globular cluster systems (black symbols). Results for the separate blue and red globular clusters are shown in the right panels. For both columns, the globular cluster effective radius is normalized to three different measures of the size of the dark matter halo: i.e., the virial radius, $R_{\rm 200}$, the radius containing half the virial mass, $R_{\rm h,dm}$, and the NFW scale radius, $R_{\rm {s,dm}}$ (top to bottom, respectively).
\label{fig:fg_dm_mdark}}
\end{figure*}
%%%%%%%%%%%%%%%%%%%%%%%%%%%%%%%%%%%%%%%%%%%%%%%%%%%%%%%%

Figure~\ref{fig:models} shows relations between $R_{200}$ and the GCS effective radius, $R_{e,{\rm gc}}$. Plots for the full GC systems are shown in the top panels, while the bottom panels show the results of fitting the blue and red GC systems separately. As described above, the symbols in the bottom panel have been color-coded by the $\Delta_{gi}$ parameter from \S\ref{data:colors}. From left to right, the four panels show the relationship between $R_{e,{\rm gc}}$ and $R_{200}$, $R_h$, $R_s$ and $\log c_{200}$, while the black, blue and red lines show the best-fit linear relations for the full, blue and red GC systems, respectively: 

%%%%%%%%%%%%%%%%%%%%%%%%%%%%%%%%%%%%%%%%%%%%%%%%%%%%%%%%
\begin{figure*}[htbp]
\centering
\includegraphics[width=0.95\textwidth]{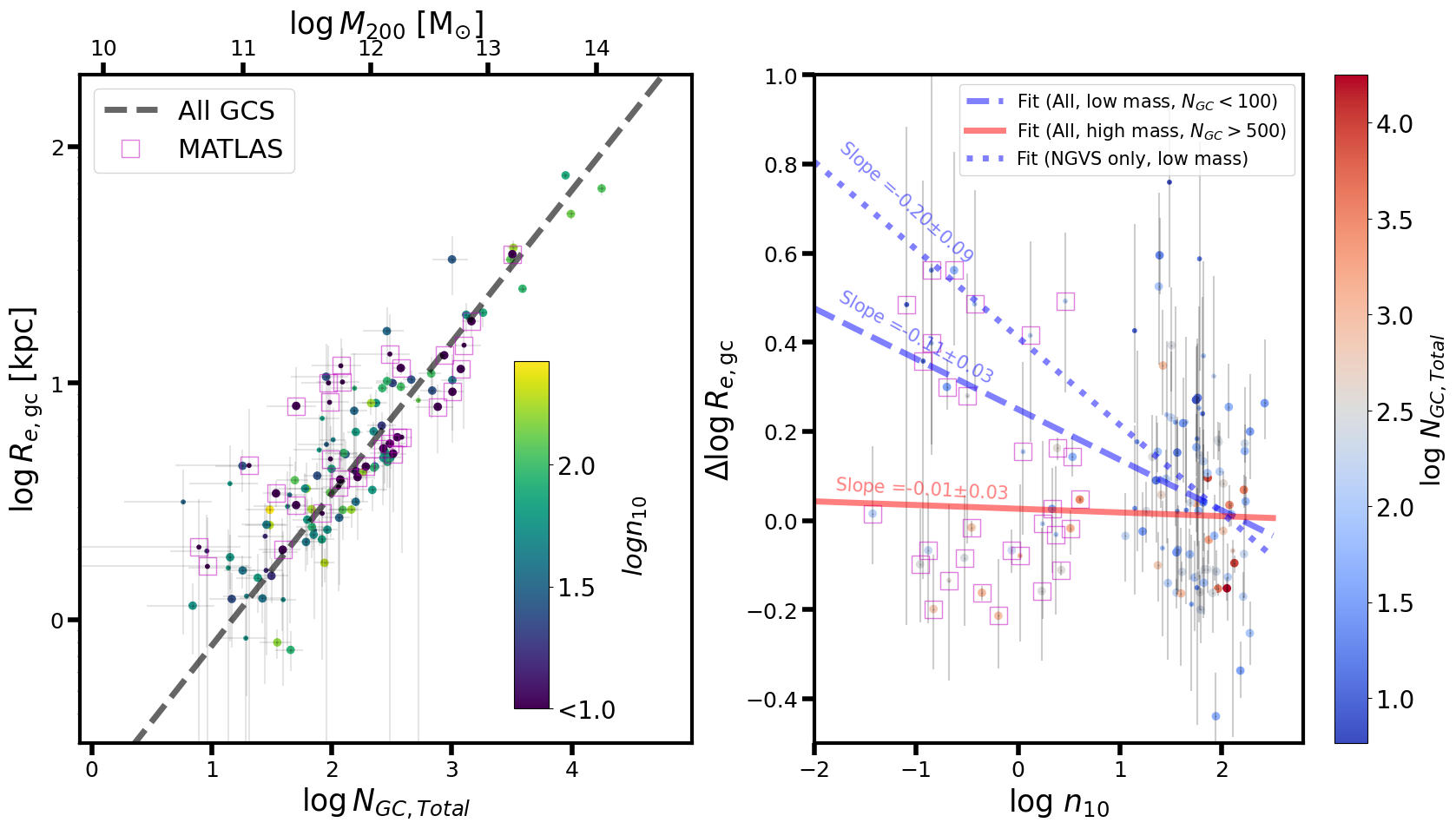}
\caption{(Left Panel) The effective radii of GC systems plotted against the total number of GCs. This is the same as in Figure~\ref{fig:numsize} but with galaxies color-coded by local density, $\log n_{\rm 10}$, and MATLAS galaxies highlighted.
(Right Panel) Residuals from the best-fit relation in the preceding panel plotted as a function of local density, $\log n_{\rm 10}$ but with galaxies color-coded by the total number of GCs. Red solid and blue dashed lines show the best-fit relation for the low mass ($N_{GC,Total} < 100$) galaxies and the high mass ($N_{GC,Total} > 500$) galaxies, respectively. The blue dotted line shows the relation found when fitting the low mass NGVS galaxies alone. MATLAS galaxies are also highlighted with open magenta squares.
\label{fig:fg_environment}}
\end{figure*}
%%%%%%%%%%%%%%%%%%%%%%%%%%%%%%%%%%%%%%%%%%%%%%%%%%%%%%%%

\footnotesize
\begin{eqnarray}
\log R_{e,{\rm gc}} & = &  (1.92\pm0.05)\log R_{\rm 200} - (3.67\pm0.12)~~~{\rm all}\\
                              & = &  (2.07\pm0.06)\log R_{\rm 200} - (4.08\pm0.16)~~~{\rm blue}\\
                              & = &  (1.41\pm0.08)\log R_{\rm 200} - (2.63\pm0.20)~~~{\rm red}
\end{eqnarray}

\normalsize

For reference, the dot-dash green lines in these panels show the relations obtained by \cite{Rei21} based on their simulated GC systems from E-MOSAICS. The shaded green band in each panel correspond to the median and 25--75th percentiles from the E-MOSAICS simulations. In the two panels at the left column, magenta dotted line and cyan dotted line show the relations from simulated GC systems presented by \citet{Che22}. They also provide the relations for {\it ex situ} and {\it in situ} GC systems which are shown with cyan dotted line and magenta dotted line, respectively. The shaded cyan and magenta bands in these panel represent $1\sigma$ confidence levels.

We begin by noting that conclusions drawn from these comparisons should be viewed with some caution for several reasons. First, our halo masses are, of course, not measured directly, but are based on an empirical relationship between number of GCs and halo mass.\footnote{The comparison presented in \S\ref{section:check} confirms that this is a reasonable assumption for galaxies with stellar masses in the upper half of our sample.} Second, the halo radii plotted in this figure were calculated under the assumption that the halos have NFW profiles, which allows us to link the halo radii to the virial masses; although this is a standard assumption, it is an assumption nonetheless. Finally, the \cite{Rei21} simulations are predictions for the GC systems of $central$ galaxies, whereas most of the galaxies in our sample are satellites. These caveats aside, we may draw a few conclusions from this comparison.

Beginning with the full GC systems, we see that the scaling relations are continuous over a range of 3.7 decades in stellar mass and have a roughly linear behaviour with surprisingly small scatter --- a direct consequence of the tight relationship between $R_{e,{\rm gc}}$ and $N_{GC, Total}$ noted in \S\ref{results:scale_galaxy} (see Figure~\ref{fig:numsize}). The $rms$ scatter around the best-fit scaling relations involving $R_{\rm 200}$, $R_h$ and $R_s$ falls in the range 0.13 to 0.20 dex. This is far smaller than the scatter in the simulations, making the tightness of the observed relations remarkable in light of the stochastic and diverse formation paths inherent in the simulations. [Note that the green bands plotted in Figure~\ref{fig:models} represent the 25--75th percentiles from the simulations, and magenta band in Figure~\ref{fig:models} shows $1\sigma$ confidence level]. Considering the overall trends, the simulations and observations match reasonably well around $R_{e,{\rm gc}} \sim 10$~kpc --- a radius corresponding to stellar and halo masses of $\sim 10^{11}M_{\odot}$ and $\sim10^{12.5}M_{\odot}$, respectively. At the high- and low-mass ends, however, the agreement is less good. At these extremes, both simulations over-predict the GC system sizes in the highest mass galaxies, and \citet{Rei21} under-predicts the sizes in the lowest mass galaxies. On the other hand, for the highest-mass galaxies, the simulations are in reasonable agreement with the measurements for the blue GC systems which, having an accretion origin in these galaxies, are more spatially extended than their red counterparts. 

As already noted from Figure~\ref{fig:sizemstar}, we find the GC effective radii to vary with galaxy stellar mass in roughly the same way as the galaxies' effective radii --- apart from a roughly constant offset of a factor of $\sim 2.5\times$, in the sense that the GC systems have more extended distributions. This similarity includes a prominent inflection at $M_p \simeq 6.5 \times10^{10}~M_{\odot}$. A standard interpretation for this bend in the galaxy relation is that (mostly) ``dry''  mergers are responsible for the high masses and steeply increasing radii of galaxies above this mass. As we have shown here (see Figures~\ref{fig:sizemstar} and \ref{fig:sizemstarcolor}), this characteristic mass also roughly divides galaxies that have strongly bimodal GC color distributions (at higher mass) from those that do not (at lower masses). Consistent with the basic picture developed to explain the size-mass relationship for galaxies, we interpret this trend as evidence that galaxies at higher mass have grown their GC systems largely through dissipationless accretion of lower mass GC systems, while galaxies below this characteristic mass have largely formed their GCs {\it in situ} --- regardless of whether the GCs are metal-poor or metal-rich. In other words, at low masses, where two GC subpopulations have similar spatial extents and colors, their origins may not be that different.

The size--$M_{200}$ relation of the red GC systems aligns well with the {\it in situ} GC systems found in \citet{Che22}. \citet{Che22} also presents this relation for {\it ex situ} GC systems, which coincides with the upper limit of the blue GC systems. Interestingly, the size relation of the {\it ex situ} simulated GC systems corresponds closely to that of the blue GC systems in massive galaxies. This may suggest that the simulation traces well the pure accreted GCs, while a significant fraction of GCs in low mass galaxies may constitute an {\it in situ} population, even though they appear relatively bluer than the galaxy's color as mentioned in \citet{Cho19}.

We also present a comparison of our results with other simulation results and previous observations in the right column of Figure~\ref{fig:models}. Consistent with our findings mentioned in Section~\ref{results:scale_galaxy}, previous observational results align with the outcomes of our study within the ranges where they were constrained by data. Notably, the simulation results of \citet{Dop23} exhibit a good match with our findings in the high mass range, but they over-predict the size of GC systems at low masses, contrary to other simulation results. The simulation result by \citet{Cre19} traces a totally different GC formation mechanism, focusing solely on accreted GC populations formed in halos that collapse before reionization, and their results are also displayed in the bottom-right of Figure~\ref{fig:models}. Their simulations produce GC systems that are much larger than the observed blue GCs. While this does not rule out such early-forming GCs, it indicates that the blue GCs need at least one other formation mechanism. 

To explore this issue in more detail, the comparisons between the effective radii of GC systems and various DM parameters are shown in Figures~\ref{fig:fg_dm_mstar} and \ref{fig:fg_dm_mdark}. 
These figures compare ratios between $R_{e,{\rm gc}}$ and the halo radii, with the panels on the left showing results for the full GC systems, and the panels on the right illustrating the trends after separation into blue and red components. 
The figures differ in the choice of independent variable, with Figure~\ref{fig:fg_dm_mstar} and Figure~\ref{fig:fg_dm_mdark} showing the trends as a function of galaxy stellar mass and halo mass, respectively. 
Relative to their dark halos, the GC systems thus do not have a constant fractional extent across the range of stellar and halo masses explored here, pointing to a somewhat more complicated formation and assembly picture. On the other hand, if we exclude the highest mass galaxies (i.e., those with $\log M_* \gtrsim 10^{11}$ or $\log M_{\rm dm} \gtrsim10^{12.5}$), then the apparent variation in fractional size of the GC system is significantly reduced, with an average value relative to the halo scale radius of $R_{e,{\rm gc}} / R_{s,{\rm dm}}\approx 0.18\pm0.03$. The fractional size of the red GC systems is more nearly constant across stellar mass, at $R_{e,{\rm gc}} / R_{s,{\rm dm}}\approx$ $0.17\pm0.01$.

\subsection{Trends with Environment}

To conclude this section, we return to the tight correlation reported in \S\ref{results} between GC system effective radius and total number of GCs. Although the scatter around this relation (Figure~\ref{fig:numsize}) is small --- with an observed $rms$ scatter of 0.21 dex which, given our measurement errors, implies an intrinsic scatter of just $\sim$ 0.1 dex --- it is interesting to note that about ten galaxies lie far above the line and about half of them are MATLAS galaxies. These galaxies are similar in many respects to the Virgo/NGVS galaxies (i.e., they have early-type morphologies, lie along the red sequence, and overlap in stellar mass), so it is natural to ask if this apparent offset is statistically significant and, if so, whether it could be related to a difference in environment. 

The left panel of Figure~\ref{fig:fg_environment} shows the $\log R_{e,{\rm gc}}$-$\log N_{GC, Total}$ relation once again but now with the MATLAS galaxies highlighted. The dashed line in this figure is same to the one in Figure~\ref{fig:numsize}.  
We have color-coded galaxies in this plot by local density, where the parameter $n_{\rm 10}$ is the density (in galaxies per arcmin$^2$) measured for all Virgo galaxies by computing the distance to the 10th nearest neighbour (and applying a geometric correction for those galaxies close to the survey boundary). For MATLAS galaxies, we estimate $n_{\rm 10}$ from the ATLAS$^{\rm 3D}$ local density parameter, $\Sigma$ \citep{Cap11}, by defining an empirical relation between these two parameters using ATLAS$^{\rm 3D}$ galaxies that fall within the NGVS footprint. Although this bootstrapping does introduce some uncertainty, the color coding in Figure~\ref{fig:fg_environment} confirms expectations that the MATLAS galaxies occupy less dense environments than do galaxies in the Virgo Cluster --- one of the richest environments in the local universe.

The situation depicted in the figure is complex. While most galaxies positioned well above the fitted line are in low-density environments, there are also galaxies in low-density environments that are close to the line (dark symbols in the left panel of Figure~\ref{fig:fg_environment}). These factors make it challenging to discern the environmental effects on the $N_{gc, Total}-R_{e,\rm gc}$ relation. To simplify the analysis, we refocus on the MATLAS sample (symbols with open squares in the left panel of Figure~\ref{fig:fg_environment}), which represents the lowest-density environments in our dataset. Within the MATLAS galaxies, we observe instances of galaxies both above and below the line, with above-the-line galaxies generally being more distant from the fitted line than below-the-line galaxies. Interestingly, most above-the-line MATLAS galaxies have small numbers of GCs with ${\rm log}N_{GC,Total} \lesssim 100$. This implies that environmental effects may vary depending on the total mass.
%%%%%%%

The right panel of Figure~\ref{fig:fg_environment} shows residuals ($\Delta \log R_{e, \rm gc}$) about the fitted relation as a function of $\log n_{\rm 10}$, but we divided samples for fitting into subgroups based on the total number of GCs. The solid red line shows the best-fit relation between $\log n_{\rm 10}$ and residuals of massive galaxies with $N_{GC, Total} > 500$, and the blue dashed line shows the fitted relation for low-mass galaxies with $N_{GC, Total} < 100$. These two datasets reveal clearly distinct trends. Effective radii of GC systems in massive galaxies do not depend on the environments. In contrast, low-mass galaxies show a clear trend for GC systems in lower-density environments to have larger effective radii at fixed $N_{GC, Total}$. Given that the $\log n_{\rm 10}$ estimates are more uncertain for MATLAS galaxies, we also show the best-fit relation for NGVS low-mass galaxies alone (the blue dotted line). In this case, the trend is even steeper, although some caution is probably appropriate given the more limited span in density. Regardless of which sample is used, though, the slope of low-mass galaxies is significant at the 2--3$\sigma$ level. In addition, this trend persists even when the low-mass cut is varied up to $N_{GC,total} \approx 300$. This trend, if confirmed by future observations, may indicate a tendency for GC systems of low-mass galaxies in low-density environments to have formed with a more extended spatial distribution. Alternatively, it may be evidence for the dynamical evolution of GC systems in the low-mass halos, with those systems in dense environments having been tidally truncated through interactions. On the other hand, GC systems in massive halos exhibit little dependence on their environments for spatial distributions, implying that massive halos may prevent tidal truncation on the dynamical evolution of GC systems in dense environments.

%%%%%%%%%%%%%%%%%%%%%%%%%%%%%%%%%%%%%%%%%%%%%%%%%%%%%%%%%%%%%%%%%%%%%%%%%%%%%%%%%%%%%%
%%%%%%%%%%%%%%%%%%%%%%%%%%%%%%%%%%%%%%%%%%%%%%%%%%%%%%%%%%%%%%%%%%%%%%%%%%%%%%%%%%%%%%
%%%
%%% DISCUSSION (Section 4)
%%%
%%%%%%%%%%%%%%%%%%%%%%%%%%%%%%%%%%%%%%%%%%%%%%%%%%%%%%%%%%%%%%%%%%%%%%%%%%%%%%%%%%%%%%
%%%%%%%%%%%%%%%%%%%%%%%%%%%%%%%%%%%%%%%%%%%%%%%%%%%%%%%%%%%%%%%%%%%%%%%%%%%%%%%%%%%%%%

\section{Summary}
\label{summary}

The size of a galaxy is a fundamental property that encodes the initial conditions of galaxy formation, as well as a Hubble time of dissipation and dynamics. The relative spatial distributions of a galaxy's gas, stars, and dark matter reflect the dominant physical processes for each component and their correspondingly different evolutionary histories. GCs, which trace the oldest stellar populations and often have the large spatial extent characteristic of stellar halos, have the potential to provide new insights into galaxy assembly histories, although spatial distributions remain one of the most poorly understand properties of GC systems.

We have combined ground- and space-based imaging for early-type galaxies in the nearby universe in order to measure the size and structure of their GC systems, and to explore correlations with the properties of their host galaxies. We have targeted early-type galaxies from the NGVS and MATLAS surveys --- two CFHT large programs undertaken with the MegaCam mosaic camera. For most of our sample galaxies, high-resolution imaging is also available from HST/ACS. Our program thus boasts several advantages over previous studies of this sort: i.e., wide-field coverage, high angular resolution in the crowded inner regions, multi-band coverage, depth, uniformity and completeness. The sample itself is both large (N=118) and spans a wide range (a factor of $\sim$2000) in stellar mass, from dwarf to giant galaxies. 

Using a homogeneous method to measure and fit the two-dimensional density distributions of GCs in our program galaxies, we find the following results:
 
\begin{enumerate}

\item The relationship between GC system size and galaxy stellar mass ($\log R_{e,{\rm gc}}$--$\log M_*$) for early-type galaxies is best characterized by a broken power law, with an $rms$ scatter of $\sim$ 0.2 dex, over the range $M_* \approx 10^{8.5}$ to $10^{11.5}~M_{\odot}$, and a ``break" at $M_p = 6.5^{+1.8}_{-1.3} \times10^{10}~M_{\odot}$. 
The relation is significantly steeper above this mass than below, with power-law slopes of $\beta = 1.30^{+0.22}_{-0.17}$ and $\alpha = 0.34\pm0.04$ in the giant and dwarf regimes, respectively. 
Above a mass of $\log M_* \sim 10$, the size-mass relation of GCs separates into two distinct branches, with the effective radii of the blue GCs exceeding those of the red GCs by a factor of $\sim$ 2.1. At low masses, the GC systems define a single, shallow relation with enhanced scatter compared to high masses.
  
\item The relation between GC system size and galaxy size ($\log R_{e,{\rm gc}}$--$\log R_e$) has a roughly linear form over the full range in mass, albeit with significant scatter ($\sigma = 0.3$ dex). Although the slope of the best-fit relation is steeper than unity ($1.22\pm0.01$), the GC systems have effective radii that are, on average, $\sim4.0$ times larger those of the galaxies. When considering the blue and red GCs separately, these factors are $\sim5.1$ and $\sim2.2$, respectively. 

\item We find a remarkably tight relation between the total number of GCs ($N_{GC, Total}$) and the GC system size ($R_{e,{\rm gc}}$), with an $rms$ scatter of just $\sigma \simeq 0.23$~dex about the best-fit linear relation. We estimate the $intrinsic$ scatter of this relation to be just $\sim 0.1$ dex. Similarly tight linear relationships apply to the blue and red GC systems. The relation defined by the blue GCs is found to be steeper than that of the red GCs --- as expected given that the blue components dominate the GC systems of high-mass galaxies.

\item We compare halo masses for our program galaxies --- estimated from $N_{GC, Total}$ using the empirical relation of \cite{Har17} --- to halo masses from \cite{Ala17} which were measured dynamically from radial velocity measurements of individual GCs. The two sets of measurements are in good agreement, which lends credence to the halo masses computed from the \cite{Har17} relation. Nevertheless, some caution is warranted as this comparison is based on just 18 galaxies, all of which have stellar masses greater than $M_* \sim 10^{10.3}~M_{\odot}$.
 
\item Compared to the simulations of \cite{Rei21} and \cite{Che22}, the GC systems in our program galaxies exhibit a relatively small scatter about the relations between $\log R_{e,{\rm gc}}$ and the various halo radii (i.e., $\log R_{\rm 200}$, $\log R_h$ and $\log R_s$). Typically, the rms scatter about these relations is just $\sim$ 0.1 dex, which would appear to present a challenge to the current generation of GC system formation models. In addition, although the predicted and observed relations ``intersect'' at intermediate sizes ($R_{e,{\rm gc}} \sim 10$~kpc) and masses ($\log M_* \sim 10^{11}$, $\log M_{\rm dm} \sim 10^{12.5}$), the agreement is rather poor at the high- and low-mass ends, where the simulations over-predict the GC system sizes in the highest mass galaxies, and show a wide range of possibilities for the lowest mass galaxies.

\item Apart from the highest-mass galaxies (i.e., those with $\log M_* \gtrsim 10^{11}$ or $\log M_{\rm dm} \gtrsim10^{12.5}$), we find the GC systems to have a roughly constant size relative to the halo scale radius, albeit with some scatter: i.e., $\approx 0.23\pm0.05$\%. The fractional size of the red GC systems are found to be nearly constant across all stellar masses, at $0.14\pm0.01$\%.

\item Our sample galaxies span a significant range in environment --- from the densest regions of Virgo's sub-clusters, to the cluster periphery, to the small groups occupied by many MATLAS galaxies. We find some evidence that the deviations from the best-fit size--number ($\log R_{e, {\rm gc}}$--$\log N_{GC, Total}$) relation are correlated with environment in the sense that GC systems of low-mass galaxies in low-density regions are, at fixed $N_{GC, Total}$, more extended than those in high-density regions. This may indicate that GC systems of low-mass galaxies in low-density environments have more extended distributions at the time of formation or, alternatively, that GC systems of low-mass galaxies in dense environments been truncated by tidal interactions. Otherwise, massive halos may reduce the effect of tidal interactions to spatial distributions of GC systems in dense environments.

\end{enumerate}

There are some obvious extensions to this work. Observations for additional galaxies, chosen to occupy an even wider range in local density, would be useful for confirming and/or characterizing possible trends with environment. High-quality measurements for additional low-mass galaxies would be helpful in understanding the size and structure of the GC systems belonging to these presumed ``building blocks'' of high-mass galaxies. Such measurements will be challenging given the small numbers of GCs associated with any single dwarf galaxy, so high-resolution, multi-band imaging will be needed to assemble GC samples that have both high completeness and low contamination. We have a follow-up study for the MATLAS dwarfs with HST observation (Marleau et al. in prep.). Dynamical mass measurements for an expanded sample of galaxies, including dwarfs, would make it possible to connect the GC system parameters to those of halo $directly$, without resorting to empirical scaling relations. On the theoretical front, the next generation of cosmological simulations will need to match these observed scaling relations, including the apparently tight correlations between GC effective radii and host galaxy properties.

\begin{acknowledgments}
S.L. acknowledges the support from the Sejong Science Fellowship Program by the National Research Foundation of Korea (NRF) grant funded by the Korea government (MSIT) (No. NRF-2021R1C1C2006790).
CL acknowledges support from the National Natural Science Foundation of China (NSFC, Grant No. 12173025, 11833005, 11933003), 111 project (No. B20019), and Key Laboratory for Particle Physics, Astrophysics and Cosmology, Ministry of Education.
C.S. acknowledges support from ANID/CONICYT through FONDECYT Postdoctoral Fellowship Project No. 3200959.
O.M. is grateful to the Swiss National Science Foundation for financial support under the grant number PZ00P2\_202104. 

IRAF was distributed by the National Optical Astronomy Observatory, which was managed by the Association of Universities for Research in Astronomy (AURA) under a cooperative agreement with the National Science Foundation.
Based on observations obtained with MegaPrime/MegaCam, a joint project of CFHT and CEA/DAPNIA, at the Canada-France-Hawaii Telescope (CFHT) which is operated by the National Research Council (NRC) of Canada, the Institut National des Science de l'Univers of the Centre National de la Recherche Scientifique (CNRS) of France, and the University of Hawaii. The observations at the Canada-France-Hawaii Telescope were performed with care and respect from the summit of Maunakea which is a significant cultural and historic site.
Funding for SDSS-III has been provided by the Alfred P. Sloan Foundation, the Participating Institutions, the National Science Foundation, and the U.S. Department of Energy Office of Science. The SDSS-III web site is http://www.sdss3.org/.
SDSS-III is managed by the Astrophysical Research Consortium for the Participating Institutions of the SDSS-III Collaboration including the University of Arizona, the Brazilian Participation Group, Brookhaven National Laboratory, Carnegie Mellon University, University of Florida, the French Participation Group, the German Participation Group, Harvard University, the Instituto de Astrofisica de Canarias, the Michigan State/Notre Dame/JINA Participation Group, Johns Hopkins University, Lawrence Berkeley National Laboratory, Max Planck Institute for Astrophysics, Max Planck Institute for Extraterrestrial Physics, New Mexico State University, New York University, Ohio State University, Pennsylvania State University, University of Portsmouth, Princeton University, the Spanish Participation Group, University of Tokyo, University of Utah, Vanderbilt University, University of Virginia, University of Washington, and Yale University.
This research is based on observations made with the NASA/ESA Hubble Space Telescope obtained from the Space Telescope Science Institute, which is operated by the Association of Universities for Research in Astronomy, Inc., under NASA contract NAS 5–26555. These observations are associated with program GO-9401.

Some of the data presented in this paper were obtained from the Mikulski Archive for Space Telescopes (MAST) at the Space Telescope Science Institute. The specific observations analyzed can be accessed via \dataset[DOI: 10.17909/yc2t-rr81]{https://doi.org/10.17909/yc2t-rr81}.

\end{acknowledgments}
\facilities{CFHT, HST/ACS} 

\appendix
\startlongtable
\begin{deluxetable}{crrc}
\tablenum{A.1}
\tablecaption{list of targets \label{tbl:galaxies}}
\tablewidth{0pt}
\tablehead{
\colhead{Name} & \colhead{RA (J2000)} &\colhead{Dec (J2000)} & \colhead{Survey} \\
\colhead{} & \colhead{[degrees] } & \colhead{[degrees] } & \colhead{}
}
\decimalcolnumbers
\startdata
NGC0524 & $21.198778$ & $9.538793$ & MATLAS \\
NGC0821 & $32.088123$ & $10.994870$ & MATLAS \\
NGC0936 & $36.906090$ & $-1.156280$ & MATLAS \\
NGC1023 & $40.100052$ & $39.063251$ & MATLAS \\
NGC2592 & $126.783669$ & $25.970339$ & MATLAS \\
NGC2685 & $133.894791$ & $58.734409$ & MATLAS \\
NGC2768 & $137.906265$ & $60.037209$ & MATLAS \\
NGC2778 & $138.101639$ & $35.027424$ & MATLAS \\
NGC2950 & $145.646317$ & $58.851219$ & MATLAS \\
NGC3098 & $150.569458$ & $24.711092$ & MATLAS \\
NGC3245 & $156.826523$ & $28.507435$ & MATLAS \\
NGC3379 & $161.956665$ & $12.581630$ & MATLAS \\
NGC3384 & $162.070404$ & $12.629300$ & MATLAS \\
NGC3457 & $163.702591$ & $17.621157$ & MATLAS \\
NGC3489 & $165.077454$ & $13.901258$ & MATLAS \\
NGC3599 & $168.862305$ & $18.110369$ & MATLAS \\
NGC3607 & $169.227737$ & $18.051809$ & MATLAS \\
NGC3608 & $169.245697$ & $18.148531$ & MATLAS \\
NGC3630 & $170.070786$ & $2.964170$ & MATLAS \\
NGC3945 & $178.307190$ & $60.675560$ & MATLAS \\
IC3032 & $182.782333$ & $14.274944$ & NGVS,ACSVCS \\
IC3065 & $183.802417$ & $14.433083$ & NGVS,ACSVCS \\
VCC200 & $184.140333$ & $13.031417$ & NGVS,ACSVCS \\
IC3101 & $184.331833$ & $11.943389$ & NGVS,ACSVCS \\
NGC4262 & $184.877426$ & $14.877717$ & NGVS,ACSVCS \\
NGC4267 & $184.938675$ & $12.798356$ & NGVS,ACSVCS \\
NGC4278 & $185.028320$ & $29.280619$ & MATLAS \\
NGC4283 & $185.086609$ & $29.310898$ & MATLAS \\
UGC7436 & $185.581458$ & $14.760722$ & NGVS,ACSVCS \\
VCC571 & $185.671417$ & $7.950306$ & NGVS,ACSVCS \\
NGC4318 & $185.680458$ & $8.198250$ & NGVS,ACSVCS \\
NGC4339 & $185.895599$ & $6.081713$ & NGVS \\
NGC4340 & $185.897141$ & $16.722195$ & NGVS,ACSVCS \\
NGC4342 & $185.912598$ & $7.053936$ & NGVS \\
NGC4350 & $185.990891$ & $16.693356$ & NGVS,ACSVCS \\
NGC4352 & $186.020833$ & $11.218333$ & NGVS,ACSVCS \\
NGC4365 & $186.117615$ & $7.317520$ & NGVS,ACSVCS \\
NGC4371 & $186.230957$ & $11.704288$ & NGVS,ACSVCS \\
NGC4374 & $186.265747$ & $12.886960$ & NGVS,ACSVCS \\
NGC4377 & $186.301285$ & $14.762218$ & NGVS,ACSVCS \\
NGC4379 & $186.311386$ & $15.607498$ & NGVS,ACSVCS \\
NGC4387 & $186.423813$ & $12.810359$ & NGVS,ACSVCS \\
IC3328 & $186.490875$ & $10.053556$ & NGVS,ACSVCS \\
NGC4406 & $186.549225$ & $12.945970$ & NGVS,ACSVCS \\
NGC4417 & $186.710938$ & $9.584117$ & NGVS,ACSVCS \\
NGC4425 & $186.805664$ & $12.734803$ & NGVS \\
NGC4429 & $186.860657$ & $11.107540$ & NGVS \\
NGC4434 & $186.902832$ & $8.154311$ & NGVS,ACSVCS \\
NGC4435 & $186.918762$ & $13.079021$ & NGVS,ACSVCS \\
NGC4442 & $187.016220$ & $9.803620$ & NGVS,ACSVCS \\
IC3383 & $187.051208$ & $10.297500$ & NGVS,ACSVCS \\
IC3381 & $187.062083$ & $11.790000$ & NGVS,ACSVCS \\
NGC4452 & $187.180417$ & $11.755000$ & NGVS,ACSVCS \\
NGC4458 & $187.239716$ & $13.241916$ & NGVS,ACSVCS \\
NGC4459 & $187.250107$ & $13.978580$ & NGVS,ACSVCS \\
NGC4461 & $187.262543$ & $13.183857$ & NGVS \\
VCC1185 & $187.347625$ & $12.450667$ & NGVS,ACSVCS \\
NGC4472 & $187.444992$ & $8.000410$ & NGVS,ACSVCS \\
NGC4473 & $187.453659$ & $13.429320$ & NGVS,ACSVCS \\
NGC4474 & $187.473099$ & $14.068673$ & NGVS,ACSVCS \\
NGC4476 & $187.496170$ & $12.348669$ & NGVS,ACSVCS \\
NGC4477 & $187.509048$ & $13.636443$ & NGVS \\
NGC4482 & $187.543292$ & $10.779472$ & NGVS,ACSVCS \\
NGC4478 & $187.572662$ & $12.328578$ & NGVS,ACSVCS \\
NGC4479 & $187.576667$ & $13.578028$ & NGVS,ACSVCS \\
NGC4483 & $187.669250$ & $9.015665$ & NGVS,ACSVCS \\
NGC4486 & $187.705933$ & $12.391100$ & NGVS,ACSVCS \\
NGC4489 & $187.717667$ & $16.758696$ & NGVS,ACSVCS \\
IC3461 & $188.011208$ & $11.890222$ & NGVS,ACSVCS \\
NGC4503 & $188.025803$ & $11.176434$ & NGVS \\
IC3468 & $188.059208$ & $10.251389$ & NGVS,ACSVCS \\
IC3470 & $188.097375$ & $11.262833$ & NGVS,ACSVCS \\
IC798 & $188.139125$ & $15.415333$ & NGVS,ACSVCS \\
NGC4515 & $188.270625$ & $16.265528$ & NGVS,ACSVCS \\
VCC1512 & $188.394000$ & $11.261889$ & NGVS,ACSVCS \\
IC3501 & $188.465083$ & $13.322583$ & NGVS,ACSVCS \\
NGC4528 & $188.525269$ & $11.321266$ & NGVS,ACSVCS \\
VCC1539 & $188.528208$ & $12.741694$ & NGVS,ACSVCS \\
IC3509 & $188.548083$ & $12.048861$ & NGVS,ACSVCS \\
NGC4550 & $188.877548$ & $12.220955$ & NGVS,ACSVCS \\
NGC4551 & $188.908249$ & $12.264010$ & NGVS,ACSVCS \\
NGC4552 & $188.916183$ & $12.556040$ & NGVS,ACSVCS \\
VCC1661 & $189.103375$ & $10.384611$ & NGVS,ACSVCS \\
NGC4564 & $189.112473$ & $11.439320$ & NGVS,ACSVCS \\
NGC4570 & $189.222504$ & $7.246663$ & NGVS,ACSVCS \\
NGC4578 & $189.377274$ & $9.555121$ & NGVS,ACSVCS \\
NGC4596 & $189.983063$ & $10.176031$ & NGVS \\
VCC1826 & $190.046833$ & $9.896083$ & NGVS,ACSVCS \\
VCC1833 & $190.081875$ & $15.935333$ & NGVS,ACSVCS \\
IC3647 & $190.221250$ & $10.476111$ & NGVS,ACSVCS \\
IC3652 & $190.243917$ & $11.184556$ & NGVS,ACSVCS \\
NGC4608 & $190.305374$ & $10.155793$ & NGVS \\
IC3653 & $190.315500$ & $11.387083$ & NGVS,ACSVCS \\
NGC4612 & $190.386490$ & $7.314782$ & NGVS,ACSVCS \\
VCC1886 & $190.414208$ & $12.247889$ & NGVS,ACSVCS \\
UGC7854 & $190.466667$ & $9.402861$ & NGVS,ACSVCS \\
NGC4621 & $190.509674$ & $11.646930$ & NGVS,ACSVCS \\
NGC4638 & $190.697632$ & $11.442459$ & NGVS,ACSVCS \\
NGC4649 & $190.916702$ & $11.552610$ & NGVS,ACSVCS \\
VCC1993 & $191.050083$ & $12.941694$ & NGVS,ACSVCS \\
NGC4660 & $191.133209$ & $11.190533$ & NGVS,ACSVCS \\
IC3735 & $191.335083$ & $13.692500$ & NGVS,ACSVCS \\
IC3773 & $191.813833$ & $10.203611$ & NGVS,ACSVCS \\
IC3779 & $191.836208$ & $12.166306$ & NGVS,ACSVCS \\
NGC4694 & $192.062881$ & $10.983624$ & NGVS \\
NGC4710 & $192.412323$ & $15.165490$ & NGVS \\
NGC4733 & $192.778259$ & $10.912103$ & NGVS \\
NGC4754 & $193.073181$ & $11.313660$ & NGVS,ACSVCS \\
NGC4762 & $193.233536$ & $11.230800$ & NGVS,ACSVCS \\
NGC5839 & $226.364471$ & $1.634633$ & MATLAS \\
NGC5846 & $226.621887$ & $1.605637$ & MATLAS \\
NGC5866 & $226.623169$ & $55.763309$ & MATLAS \\
PGC058114 & $246.517838$ & $2.906550$ & MATLAS \\
NGC6548 & $271.496826$ & $18.587217$ & MATLAS \\
NGC7280 & $336.614899$ & $16.148266$ & MATLAS \\
NGC7332 & $339.352173$ & $23.798351$ & MATLAS \\
NGC7457 & $345.249725$ & $30.144892$ & MATLAS \\
NGC7454 & $345.277130$ & $16.388371$ & MATLAS \\
\enddata
% \tablecomments{This table ``hides'' the third column in the \latex\ when compiled.
% The Distance is also centered on the decimals.  Note that when using decimal
% alignment you need to include the {\tt\string\decimals} command before
% {\tt\string\startdata} and all of the values in that column have to have a
% space before the next ampersand.}
\end{deluxetable}

\end{document}